\documentclass[lettersize,journal]{IEEEtran}

\usepackage{amsmath,amsfonts}
\usepackage[linesnumbered,ruled,vlined]{algorithm2e}
\usepackage{array}
\usepackage{orcidlink}
\usepackage[caption=false,font=normalsize,labelfont=sf,textfont=sf]{subfig}
\usepackage{textcomp}
\usepackage{stfloats}
\usepackage{url}
\usepackage{verbatim}
\usepackage{graphicx}
\usepackage{cite}
\usepackage{booktabs}
\usepackage{multirow}
\usepackage{stfloats}
\usepackage{amsmath}
\usepackage{color}
\newtheorem{theorem}{Theorem}

\begin{document}

\title{Incentive-Driven Task Offloading and Collaborative Computing in Device-Assisted MEC Networks}

\author{Yang Li$^{\orcidlink{0009-0007-1601-633X}}$, Xing Zhang$^{\orcidlink{0000-0003-4345-6166}}$,~\IEEEmembership{Senior Member,~IEEE,} Bo Lei$^{\orcidlink{0000-0002-6301-048X}}$, Qianying Zhao$^{\orcidlink{0009-0005-9852-827X}}$, Min Wei$^{\orcidlink{0009-0006-1661-5941}}$, Zheyan Qu$^{\orcidlink{0009-0009-2679-4359}}$, and Wenbo Wang$^{\orcidlink{0000-0002-0911-3189}}$,~\IEEEmembership{Senior Member,~IEEE}
\thanks{Manuscript received 4 November 2024; accepted 25 November 2024. This work is supported by the National Science Foundation of China under Grant 62071063, 62271062, and by the BUPT Excellent Ph.D. Students Foundation under Grant CX20241066. (Corresponding author: Xing Zhang.)}
\thanks{Y. Li, X. Zhang, Z. Qu, and W. Wang are with the School of Information and Communications Engineering, Beijing University of Posts and Telecommunications, Beijing 100876, China (e-mail: ly209991@bupt.edu.cn; zhangx@ieee.org; zheyanqu@bupt.edu.cn; wbwang@bupt.edu.cn).}
\thanks{B. Lei, Q. Zhao, and M. Wei are with Beijing Branch of China Telecom Co., Ltd., Beijing 100032, China (e-mail: leibo@chinatelecom.cn; zhaoqy50@chinatelecom.cn; weim6@chinatelecom.cn).}
\thanks{Copyright (c) 2024 IEEE. Personal use of this material is permitted. However, permission to use this material for any other purposes must be obtained from the IEEE by sending a request to pubs-permissions@ieee.org.}
}

\markboth{IEEE INTERNET OF THINGS JOURNAL}%
\IEEEaftertitletext{\vspace{0\baselineskip}}

\maketitle

\begin{abstract}
Edge computing (EC), positioned near end devices, holds significant potential for delivering low-latency, energy-efficient, and secure services. This makes it a crucial component of the Internet of Things (IoT). However, the increasing number of IoT devices and emerging services place tremendous pressure on edge servers (ESs). To better handle dynamically arriving heterogeneous tasks, ESs and IoT devices with idle resources can collaborate in processing tasks. Considering the selfishness and heterogeneity of IoT devices and ESs, we propose an incentive-driven multi-level task allocation framework. Specifically, we categorize IoT devices into task IoT devices (TDs), which generate tasks, and auxiliary IoT devices (ADs), which have idle resources. We use a bargaining game to determine the initial offloading decision and the payment fee for each TD, as well as a double auction to incentivize ADs to participate in task processing. Additionally, we develop a priority-based inter-cell task scheduling algorithm to address the uneven distribution of user tasks across different cells. Finally, we theoretically analyze the performance of the proposed framework. Simulation results demonstrate that our proposed framework outperforms benchmark methods.
\end{abstract}


\section{Introduction}\label{I}

\subsection{Research Background and Motivation}\label{I-A}
\IEEEPARstart{T}{he} rapid development of the Internet of Things (IoT) and 5G technologies has led to an increased demand for computationally intensive and latency-sensitive tasks, such as face recognition, Internet of Vehicles, and AR/VR. Resource-constrained IoT devices cannot process these tasks locally. Traditional cloud computing introduces long backhaul link latency, failing to provide real-time computation \cite{1}. Therefore, mobile edge computing (MEC) has emerged as a promising paradigm to support various 5G service scenarios, particularly low-latency services \cite{2}. However, due to the limited computational capabilities of edge servers (ESs) and the increasing number of IoT devices, meeting the demand for numerous latency-sensitive, computationally intensive tasks with resource-limited ESs remains a significant challenge. Additionally, cost constraints prevent the continuous expansion of ES resources \cite{3}. In this context, device-assisted edge computing has gained increased scholarly attention \cite{4,5,6,7,8,9,10,11,+1,+2}. Leveraging the diversity among IoT devices and incorporating those with idle resources as resource providers to assist ESs in processing tasks can enhance the computing capacity of edge systems while improving the utilization of end-device resources \cite{4}.

Current research on device-assisted MEC networks primarily focuses on computation offloading and content caching \cite{4,5,6,7,8,9,10,11,+1,+2}. In device-assisted computation offloading, users can offload tasks to edge servers (ESs) or nearby auxiliary devices. Offloading tasks to nearby auxiliary devices over device-to-device (D2D) links can reduce the load on the cellular network infrastructure and free up cellular bandwidth for other uses \cite{4,5,6,7}. The main strategy of current device-assisted MEC caching approaches is to prioritize content placement and delivery from other end devices through D2D communication links. This is followed by content placement and delivery from the MEC server, and lastly, from the central cloud \cite{8,9,10,11,+1,+2}. Generally, content caching has two main aspects: content placement and content delivery. Content placement studies focus on designing methods for optimally storing (placing) content files in the caches of ESs and IoT devices. Conversely, content delivery studies concentrate on transmitting the requested files to end devices.

Overall, device-assisted MEC research has made significant progress in computational offloading and content caching. Nevertheless, device-assisted MEC is still an emerging research area, and current state-of-the-art research has significant limitations. For example, most studies consider simple schemes, usually involving only one ES. Collaboration among different ESs is intriguing but rarely addressed. Existing collaboration schemes among ESs are excessively complex, preventing their application in realistic scenarios. Additionally, since assisting with processing tasks consumes energy, auxiliary IoT devices are reluctant to help ESs without incentives. However, research on incentive mechanisms in device-assisted edge computing is limited. To address this, we propose an incentive-driven multi-level task allocation scheme to encourage cooperation among participants for efficient task processing. Specifically, our research addresses the following challenges:
\begin{enumerate}
\item{Device-assisted edge computing networks involve various participants with differing intentions. Without suitable incentives, efficient collaboration among participants may be hindered.}
\item{Varied task requirements, execution mode choices, and resource constraints result in a combinatorial optimization and mixed-integer problem, which is NP-hard.}
\item{In a multi-ES scenario, collaboration among ESs can enhance system performance but also introduces additional computational complexity, requiring a balance between algorithm performance and complexity.}
\end{enumerate}

\subsection{Related Work}\label{I-B}
\begin{table*}[b]
\centering
\caption{Overview of the related work.}
\label{Overview of the related work}
\resizebox{0.9\textwidth}{!}{%
\begin{tabular}{llll}

\multicolumn{4}{c}{\textit{\textbf{Incentive Mechanism for Resource Allocation}}}\vspace{0.1cm}                                                                                                                                                                                                                                                                                                                                                                                                                             \\ \midrule[1.5pt]
{\textbf{Research}}        & \textbf{\begin{tabular}[c]{@{}l@{}}Incentive \\ Mechanism\end{tabular}}                                                                                     & \textbf{Participants}                                                                                                                                        & \textbf{Objective}                                                                                                                                                        \\ \midrule[1.5pt]
{\cite{12,+3,13,14,15,16}} & Stackelberg game                                                                                            & ESs and task IoT devices (TDs)                                                                                                                               & \begin{tabular}[c]{@{}l@{}}Design service pricing for ESs and maximize \\ the utility of both ESs and TDs.\end{tabular}                                                   \\ \hline
{\cite{17,+4,18,19 }}      & Auction theory                                                                                              & ESs and TDs                                                                                                                                                  & \begin{tabular}[c]{@{}l@{}}Ascertain resource allocation and transaction\\ prices in MEC networks.\end{tabular}                                                           \\ \hline
{\cite{20,+5,21,22,23}}    & Contract theory                                                                                             & ESs and auxiliary IoT devices (ADs)                                                                                                                          & \begin{tabular}[c]{@{}l@{}}Motivate idle devices to provide resources \\ and maximize the utility of ESs.\end{tabular}                                                    \\ \hline
{\cite{25,24,26,+6}}       & Bargaining game                                                                                             & ESs and TDs                                                                                                                                                  & \begin{tabular}[c]{@{}l@{}}Ascertain resource allocation and maximize \\ the utility of both ESs and TDs.\end{tabular}                                                    \\ \hline
{Our study}                & \begin{tabular}[c]{@{}l@{}}Bargaining game, \\ priority-based \\scheduling, \\ and double action\end{tabular} & ESs, TDs, and ADs                                                                                                                                            & \begin{tabular}[c]{@{}l@{}}Motivate horizontal collaboration among ESs \\ and vertical collaboration between ESs and ADs, \\ and maximize the system utility.\end{tabular} \\ \midrule[1.5pt]
\\
\multicolumn{4}{c}{\textit{\textbf{Task Allocation in Device-Assisted MEC}}}    \vspace{0.1cm}                                                                                                                                                                                                                                                                                                                                                                                                                                  \\ \midrule[1.5pt]
{\textbf{Research}}        & \textbf{Scenario}                                                                                           & \textbf{Technical method}                                                                                                                                    & \textbf{Objective}                                                                                                                                                        \\ \midrule[1.5pt]
{\cite{27}}                & Single ES                                                                                                 & \begin{tabular}[c]{@{}l@{}}Lyapunov optimization and \\ variable substitution technique\end{tabular}                                                         & \begin{tabular}[c]{@{}l@{}}Minimize the network-wide response latency \\ and energy consumption.\end{tabular}                                                             \\ \hline
{\cite{4}}                 & Single ES                                                                                                 & \begin{tabular}[c]{@{}l@{}}Deep reinforcement learning \\ and graph theory\end{tabular}                                                                      & \begin{tabular}[c]{@{}l@{}}Minimize the maximum processing latency for \\ all tasks.\end{tabular}                                                                         \\ \hline
{\cite{5}}                 & Single ES                                                                                                 & Iteration-based programme                                                                                                                                    & \begin{tabular}[c]{@{}l@{}}Minimize the sum of task execution latency of \\ all the devices.\end{tabular}                                                                 \\ \hline
{\cite{+7}}                & Single ES                                                                                                 & Block descent and potential game                                                                                                                             & \begin{tabular}[c]{@{}l@{}}Improve resource utilization and reduce computing \\ latency.\end{tabular}                                                                      \\ \hline
{\cite{7}}                 & Multiple ESs                                                                                                  & Multi-armed Bandit                                                                                                                                           & \begin{tabular}[c]{@{}l@{}}Minimize the weighted sum of task latency and \\ service migration costs.\end{tabular}                                                         \\ \hline
{\cite{6}}                 & Multiple ESs                                                                                                  & Game theory                                                                                                                                                  & \begin{tabular}[c]{@{}l@{}}Minimize the overall offloading cost in terms of \\ processing delay and energy consumption.\end{tabular}                                      \\ \hline
{\cite{31}}                & Multiple ESs                                                                                                  & Graph theory                                                                                                                                                 & Maximize the system utility.                                                                                                                                               \\ \hline
{Our study}                & Multiple ESs                                                                                                  & \begin{tabular}[c]{@{}l@{}}Multi-level decision-making, \\ which integrates bargaining game, \\ priority-based scheduling, \\ and double action\end{tabular} & \begin{tabular}[c]{@{}l@{}}Motivate horizontal collaboration among ESs and \\ vertical collaboration between ESs and ADs, and \\ maximize the system utility.\end{tabular} \\ \midrule[1.5pt]
\end{tabular}%
}
\end{table*}
\textit{1) Incentive Mechanism for Resource Allocation:} The incentive mechanisms for resource allocation in current research can be broadly categorized into Stackelberg game-based approaches \cite{12,+3,13,14,15,16}, auction-based approaches \cite{17,+4,18,19 }, contract-theory-based approaches \cite{20,+5,21,22,23}, and bargaining-based approaches \cite{25,24,26,+6}.

The Stackelberg game divides players into leaders and followers. The leader acts first, and the followers choose the optimal response based on the leader's strategy. The leader's optimal strategy combines with the followers' optimal responses to form a Stackelberg equilibrium. Zhou \textit{et al.} \cite{12} used the Stackelberg game to model interactions between edge service providers (ESPs) and mobile users (MUs). They used backward induction to analyze the game problem, aiming to maximize the utility of both ESPs and MUs. Fan \textit{et al.} \cite{+3} proposed a two-stage pricing scheme for MEC network slicing. The authors used a Stackelberg game to model the interactions between MEC network service providers (MEC-NSPs) and user devices (UDs). In the first stage, they optimized resource allocation, and in the second stage, they determined the slice pricing. Similarly, the studies in \cite{13,14,15,16} used the Stackelberg game to design service pricing for resource providers.

Auction theories offer an efficient way to create a market where resource seekers bid for one or more resources, effectively providing a theoretical basis for resource allocation. Wang \textit{et al.} \cite{17} designed an online profit-maximizing multi-round auction mechanism for trading resources between edge clouds (sellers) and mobile devices (buyers) in competitive environments. Habiba \textit{et al.} \cite{+4} proposed a repeated auction model for dynamic resource allocation in MEC networks. The studies in \cite{18} and \cite{19} employ a double auction mechanism to ascertain matching outcomes and transaction prices for IoT devices and edge servers.

Contract theory offers valuable tools for designing incentive mechanisms. It involves two parties: agents (e.g., user equipment (UE)) and trustors (e.g., base stations (BSs) or ESs). This mechanism avoids the limitations of long convergence times and the need for multiple information exchanges between agents and trustors, which can result from iteration-based schemes like the Stackelberg game and auction mechanism. However, it must address the challenges posed by information asymmetry. Chen \textit{et al.} \cite{20} proposed a signaling-based incentive mechanism using contract theory to address information asymmetry in the D2D computation offloading problem. The authors in \cite{+5} proposed a contract incentive strategy based on deep reinforcement learning for a wireless-powered, UAV-assisted backscattering MEC system. Their objective was to maximize the long-term utility of hotspots while maintaining the stability of energy queues. Many studies have used contract theory to design incentive mechanisms, primarily to motivate idle devices to provide resources \cite{21,22,23}.

The bargaining game is a cooperative game with the features of incentive, self-enforcement, and satisfaction for all players \cite{25}. It guarantees fairness and Pareto-optimality of the outcome and captures the potential of coordination among bargainers \cite{24}. Shi \textit{et al.} \cite{26} investigated the service deployment problem in fog computing using the bargaining game concept to enhance service providers' revenues. Meskar \textit{et al.} \cite{+6} proposed a fair resource allocation mechanism for heterogeneous servers in MEC networks. They employed the Kalai-Smorodinsky bargaining solution to ensure fairness properties, including envy-freeness, Pareto optimality, strategy-proofness, and sharing incentives.

However, the previous studies mainly focus on the interactions between ESs and task IoT devices (TDs), or ESs and auxiliary IoT devices (ADs), by designing incentives to motivate cooperation between two parties with conflicting interests. This paper considers a scenario involving three types of participants: ESs, TDs, and ADs. In this scenario, different ESs share common interests, while conflicts of interest exist among TDs, among ADs, and between TDs, ADs, and ESs. Therefore, none of the previously proposed incentive mechanisms can be directly applied. Unlike previous work, this paper thoroughly analyzes the interactions among the three types of participants and proposes a multi-level decision-making framework. This framework integrates bargaining games, priority-based scheduling, and double auctions, effectively leveraging the cooperative and competitive relationships among the participants to maximize system utility.

\textit{2) Task Allocation in Device-Assisted MEC:} Researchers have done much work on task allocation in device-assisted MEC networks for single ES and multiple ES scenarios.

For scenarios involving a single ES, Peng \textit{et al.} \cite{27} investigated the joint optimization of collaborative partial offloading, transmission scheduling, and task allocation for device-assisted MEC networks. They proposed an online resource coordination and allocation scheme (ORCAS). Our previous work \cite{4} proposed a joint task partitioning and parallel scheduling framework. This framework first divides computationally intensive tasks into multiple subtasks. Then, it schedules these subtasks to be processed in parallel on multiple ADs and the ES, enabling collaboration among the ADs and the ES. The studies in \cite{5} and \cite{+7} both focused on minimizing the overall task execution delay within single-cell, device-assisted MEC networks. However, a single ES has limited computational resources. Although some studies \cite{28,29,30,+8,+9} have investigated cloud cooperation to compensate for the limited resources of a single ES, this approach introduces significant backhaul delay. Leveraging collaboration among multiple ESs is an effective way to address this challenge.

For scenarios involving multiple ESs, Dai \textit{et al.} \cite{7} proposed a collaborative offloading framework that integrates migration cost and offloading willingness in a device-assisted MEC network to minimize the system cost. However, instead of leveraging collaboration among ESs, this work uses a learning-based approach to prompt IoT devices to select the best computing node. Yang \textit{et al.} \cite{6} proposed a game-theory-based offloading approach that treats the computation offloading process in a device-assisted MEC network as a competitive game for resources, thereby minimizing system overhead and the execution cost of individual tasks. However, as the number of ESs and IoT devices increases, the long convergence time of the game-theoretic algorithm can prevent its application in real-world scenarios. Hou \textit{et al.} \cite{31} proposed an online task allocation algorithm to optimize the task allocation policy in a device-assisted MEC network, maximizing the system utility. However, the proposed scheme only considers serially arriving tasks, meaning only one task's processing position can be determined at a time. In real IoT systems, tasks from multiple users often arrive simultaneously, requiring the processing locations of multiple tasks to be determined concurrently.

To our knowledge, few studies consider collaboration among multiple cells in device-assisted MEC networks. As the number of cells increases, the linear growth in the number of TDs, ADs, and ESs complicates problem-solving. The most relevant studies \cite{6}, \cite{31} proposed schemes that were highly sensitive to the number of cells, rendering them unsuitable for online use in device-assisted MEC networks with a large number of cells. The challenge in implementing online assignment of concurrent task requests in multi-cell device-assisted MEC networks lies in optimizing system utility while considering algorithm complexity. Unlike previous work, this paper introduces a multi-level decision-making framework for managing concurrent task requests in multi-cell device-assisted MEC networks, enabling efficient collaboration among ESs, ADs, and TDs. Moreover, this framework allows the system to handle varying task complexity and resource availability, remains insensitive to the number of cells. It also adapts algorithmic complexity based on user request distribution by either concluding early or bypassing the second level of decision-making, demonstrating high flexibility and scalability.

Table \ref{Overview of the related work} summarizes the closely related studies on incentive mechanisms for resource allocation and task allocation in device-assisted MEC networks, comparing these works with ours.

\subsection{Contribution and Organization}\label{I-C}
\begin{figure}[h]
\centerline{\includegraphics[width=0.5\textwidth]{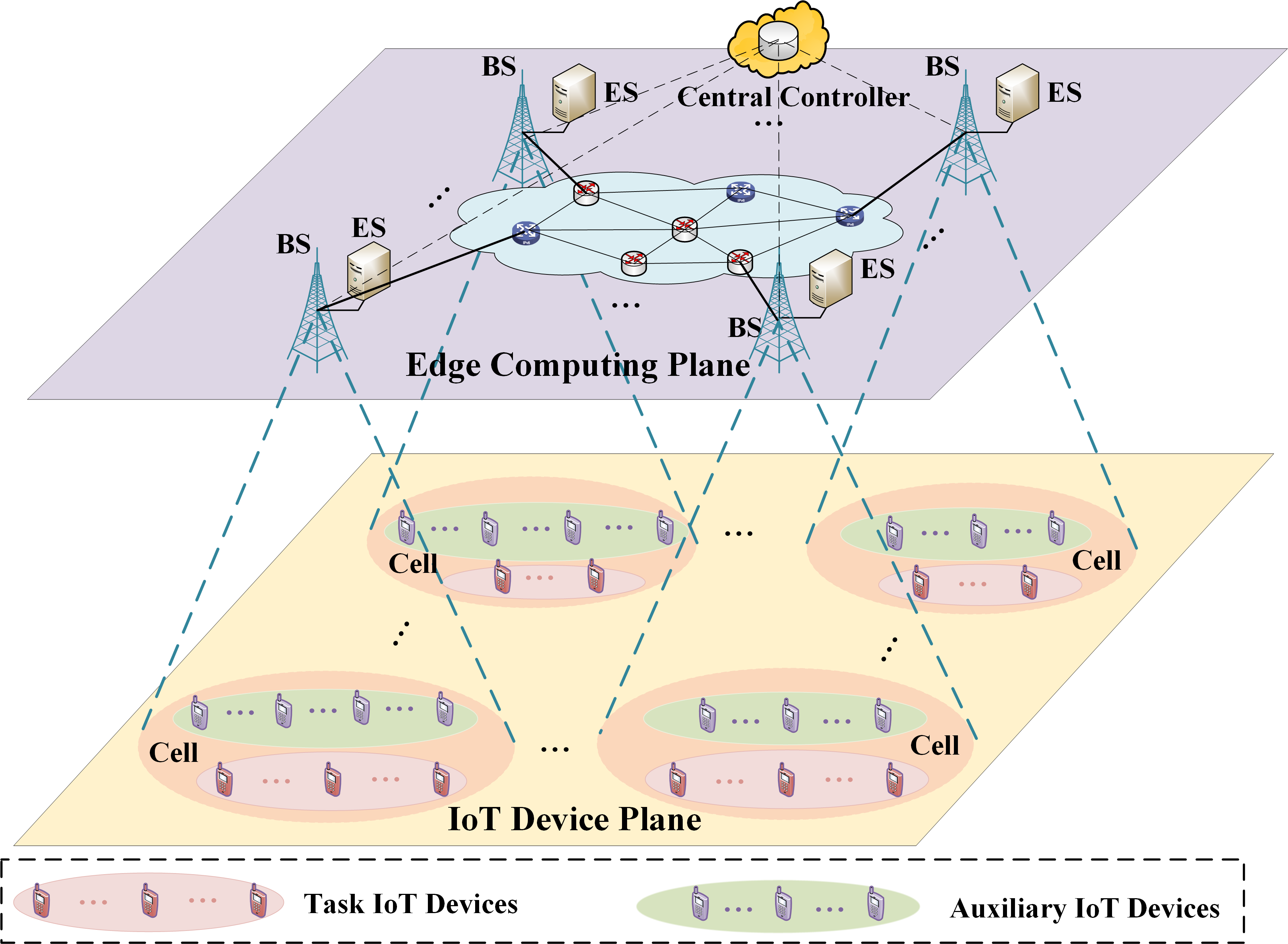}}
    \caption{System model of the device-assisted MEC network.}
\label{system}
\vspace{-0.5cm}
\end{figure}

In this paper, we investigate the device-assisted MEC network depicted in Fig. \ref{system}. To promote effective collaboration among various system participants and achieve efficient task processing, we propose an incentive-driven multi-level task allocation scheme. The main contributions of this paper are summarized as follows:
\begin{enumerate}
\item{We propose a hierarchical task assignment scheme with collaborative computing for device-assisted MEC networks. This multi-level task allocation strategy enables scalable and low-complexity scheduling decisions for concurrent heterogeneous tasks.}
\item{At the first level of decision-making, we model the interaction between each TD and the ES in each cell as a bargaining game to maximize their utility. Through bargaining, the offloading decision and payment fee for each TD are determined. This mechanism ensures fairness and Pareto optimality in task offloading between TDs and ESs.}
\item{If some ESs are overloaded after the first level of decision-making, they will collaborate with underloaded ESs to reassign tasks. We design a priority-based task filtering scheme and task scheduling algorithm for the second level of decision-making.}
\item{If all ESs' resources are still insufficient, we introduce a double auction mechanism to incentivize ADs to assist ESs in the third level of decision-making.}
\item{Numerical simulations show that the proposed scheme outperforms the benchmark methods in system utility, task offloading ratio, algorithm execution delay, and load balancing capability.}
\end{enumerate}

The rest of the paper is organized as follows. Section II introduces the system model and problem formulation. Section III presents the proposed multi-level task allocation algorithm. Section IV presents the theoretical analysis of the proposed framework. Section V discusses the simulation results and performance analysis. Finally, Section VI concludes the paper. The code is available at https://github.com/CPNGroup/Multi\_Level\_EC.

\section{System model and problem formulation}
\subsection{System Overview}\label{II-A}
As illustrated in Fig. \ref{system}, we consider an IoT network comprising $M$ cells. Each cell is covered by a base station (BS) integrated with an ES, and all ESs are centrally controlled by an edge service provider (ESP) through a controller. For convenience, we use the same notation $\mathcal{M}=\{1,2,...,M\}$ to represent the sets of cells, BSs, and ESs. Moreover, several IoT devices are distributed in each cell. Similar to previous studies \cite{4,5}, we categorize IoT devices into two types: task IoT devices (TDs) that generate tasks and can enhance their utility through computation offloading, and auxiliary IoT devices (ADs) with idle resources that can assist the ESs in their respective cells to process tasks. Assuming that in the $m$th cell, there are $N_m$ TDs denoted as $\mathcal{N}_m=\{n_1^m,n_2^m,...,n_{N_m}^m\}$ and $K_m$ ADs denoted as $\mathcal{K}_m=\{k_1^m,k_2^m,...,k_{K_m}^m\}$. Consequently, the sets of all TDs and ADs in the network can be denoted as $\boldsymbol{\mathcal{N}}=\{\mathcal{N}_1,...,\mathcal{N}_M\}$ and $\boldsymbol{\mathcal{K}}=\{\mathcal{K}_1,...,\mathcal{K}_M\}$, respectively. Additionally, TDs, ADs and the ESP are typical interest-driven entities with conflicts of interest. TDs obtain edge computing services from the ESP and pay computation fees, while ADs assist the ESP in processing tasks and receive rewards from the ESP. The ESP earns revenue from providing services to TDs and can also profit by outsourcing tasks to ADs, acting as a middleman to capture a margin.

We assume that different cells reuse the same wireless resources, and IoT devices associated with the same BS utilize orthogonal spectrums for data transmission. Additionally, we assume that each IoT device within a cell occupies one sub-band, and the sub-band allocation among IoT devices is randomized. Consistent with \cite{31}, we consider a quasi-static network scenario, where each IoT device's channel remains unchanged until computation offloading is completed. Table \ref{tab1} lists the main notations used in this paper.

\begin{table}[]
\centering
\caption{Summary of system notations
}
\label{tab1}
\resizebox{\columnwidth}{!}{%
\begin{tabular}{ll}
\hline
\textbf{Notation}         & \textbf{Description}                                                                                                                         \\ \hline
$\mathcal{M}$             & The set of BSs/cells/ESs                                                                                                                     \\
$\mathcal{N}$             & The set of task IoT devices                                                                                                                  \\
$\mathcal{K}$             & The set of auxiliary IoT devices                                                                                                             \\
$n_i^m$                   & The $i$th TD in the $m$th cell                                                                                                               \\
$k_j^m$                   & The $j$th AD in the $m$th cell                                                                                                               \\
$m$                       & The $m$th BS/cell/ES                                                                                                                         \\
$R_i^j$                   & \begin{tabular}[c]{@{}l@{}}Data transfer rate between sender $i$ and \\ receiver $j$\end{tabular}                                            \\
$H_{n_i^m}$               & The task generated by TD $n_i^m$                                                                                                             \\
$L_{n_i^m}$               & The size of $H_{n_i^m}$                                                                                                                      \\
$C_{n_i^m}$               & CPU cycles required to process $H_{n_i^m}$                                                                                                   \\
$t_{n_i^m}^{max}$         & Maximum delay tolerated for task $H_{n_i^m}$                                                                                                 \\
$V_{n_i^m}$               & \begin{tabular}[c]{@{}l@{}}Value to TD $n_i^m$ when $H_{n_i^m}$ is \\ successfully processed\end{tabular}                                    \\
$x_{n_i^m}$               & Offloading decision for TD $n_i^m$                                                                                                           \\
$y_{n_i^m}$               & Execution mode of $H_{n_i^m}$                                                                                                                \\
$z_{n_i^m}$               & Processing position of $H_{n_i^m}$                                                                                                           \\
$u_{n_i^m}$               & Utility of TD $n_i^m$                                                                                                                        \\
$u_{esp}$                 & Utility of the ESP                                                                                                                           \\
$u_{k_j^m}$               & Utility of AD $k_j^m$                                                                                                                        \\
$\gamma$                  & Cost per unit of energy consumption                                                                                                          \\
$F_i$                     & Total available resources for computing node $i$                                                                                             \\
$f_j^{n_i^m}$             & \begin{tabular}[c]{@{}l@{}}Computational resources that node $j$ needs to \\ allocate for task $H_{n_i^m}$\end{tabular}                      \\
$t_{a,b}^{n_i^m,tran}$    & \begin{tabular}[c]{@{}l@{}}The transmission delay of task $H_{n_i^m}$ from\\ computing node $a$ to computing node $b$.\end{tabular}          \\
$cost_{a,b}^{n_i^m,tran}$ & \begin{tabular}[c]{@{}l@{}}The cost of transferring task $H_{n_i^m}$ from\\ computing node $a$ to computing node $b$.\end{tabular}           \\
$cost_{a}^{n_i^m,comp}$   & \begin{tabular}[c]{@{}l@{}}The cost of processing task $H_{n_i^m}$ on\\ computing node $a$.\end{tabular}                                     \\
$u_m^{n_i^m}$             & The gain that ES $m$ can obtain from TD $n_i^m$.                                                                                             \\
$u_{esp}^{n_i^m,a}$       & \begin{tabular}[c]{@{}l@{}}The utility gained by the ESP when assigning \\ task $H_{n_i^m}$ to compute node $a$ for processing.\end{tabular} \\
$\alpha_{n_i^m}$          & The fee that TD $n_i^m$ needs to pay to the ESP                                                                                              \\
$\alpha_{k_j^m}$          & The fee that the ESP needs to pay to AD $k_j^m$                                                                                              \\ \hline
\end{tabular}%
}
\end{table}

\begin{figure*}[b]
\centerline{\includegraphics[width=\textwidth]{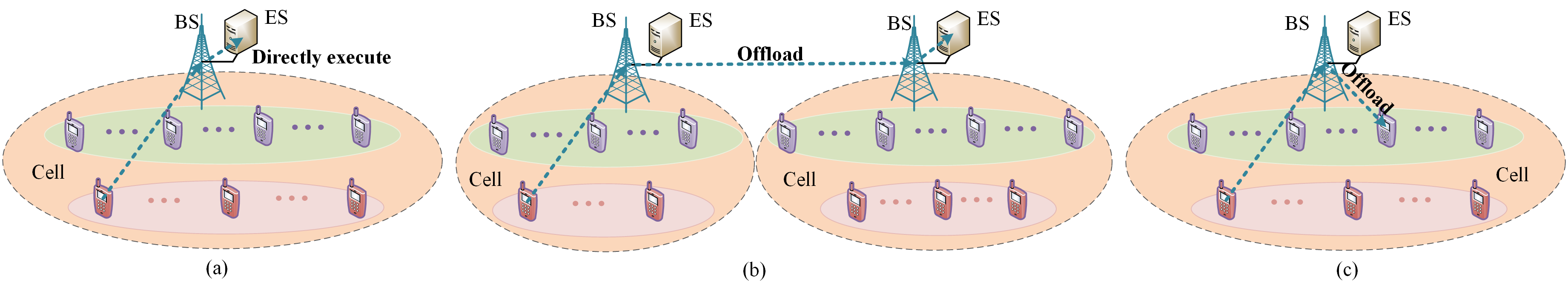}}
    \caption{Execution modes. (a) Primary ES execution. (b) Neighboring ES execution. (c) AD execution.}
\label{mode}
\end{figure*}

\subsection{Task Model}\label{II-B}
The task of TD $n_i^m$, $\forall n_i^m \in \boldsymbol{\mathcal{N}}$ is denoted as $H_{n_i^m} = \{L_{n_i^m},C_{n_i^m},t_{n_i^m}^{max}\}$. Here, $L_{n_i^m}$ denotes the size of $H_{n_i^m}$, $C_{n_i^m}$ denotes the number of CPU cycles required to complete $H_{n_i^m}$, and $t_{n_i^m}^{max}$ denotes the maximum tolerable delay for $H_{n_i^m}$. The parameters $\{L_{n_i^m}, C_{n_i^m},t_{n_i^m}^{max}\}$ can be obtained from the task profiler \cite{32}. We assume these tasks are atomic, and $x_{n_i^m}$ denotes the offloading decision of TD $n_i^m$, where $x_{n_i^m}=1$ indicates that TD $n_i^m$ offloads its task, and $x_{n_i^m}=0$ indicates that TD $n_i^m$ processes its task locally.

\subsection{Execution Model}\label{II-C}
To fully utilize the ubiquitously distributed computational resources, collaboration can be applied among ESs in different cells (horizontal collaboration) and between the ES and ADs within each cell (vertical collaboration). We define the ES of each cell as the primary ES for all TDs in that cell, i.e., ES $m$ is the primary ES for TD $n_i^m$, $\forall n_i^m  \in \mathcal{N}_m$. Additionally, we denote ES $m^{\prime}$, $\forall m^{\prime} \in \mathcal{M} \backslash \{m\}$ as the neighboring ES for TD $n_i^m$, $\forall n_i^m  \in \mathcal{N}_m$. Based on the previous definitions, TD $n_i^m$ has four execution modes for task processing: 1) local execution, 2) primary ES execution, 3) neighboring ES execution, and 4) AD execution. We define $y_{n_i^m}$ as the execution mode of the task $H_{n_i^m}$, where $y_{n_i^m}\in \{loc, pe, ne, ad\}$ corresponds to the four execution modes described earlier, respectively. Combining with the previous definition of $x_{n_i^m}$, we can easily obtain $y_{n_i^m}=loc$ when $x_{n_i^m}= 0$, and $ y_{n_i^m}\in \{pe, ne, ad\}$ when $x_{n_i^m}= 1$. The cost of relevant participants in the above four different execution modes is analyzed as follows:

\textit{\textbf{1) Local execution mode:}} When TD $n_i^m$ processes its task locally, dynamic voltage and frequency scaling techniques can be utilized to control the energy consumption \cite{33}. Considering the maximum tolerable delay $t_{n_i^m}^{max}$ for task $H_{n_i^m}$, the CPU frequency $ f_{n_i^m}^{n_i^m}$ used by TD $n_i^m$ to process $H_{n_i^m}$ needs to satisfy $ f_{n_i^m}^{n_i^m}\geq C_{n_i^m}/ t_{n_i^m}^{max} $. According to \cite{34}, the energy consumption is proportional to the square of the computation frequency. We assume that under the incentive of individual rationality, all computing nodes process tasks at the lowest CPU frequency that satisfies the task demand. Therefore, the CPU frequency of TD $n_i^m$ for local computation is $ f_{n_i^m}^{n_i^m}=C_{n_i^m}/ t_{n_i^m}^{max} $. At this point, the cost of TD $n_i^m$ can be calculated as described in \cite{34}:
\begin{equation}
\label{eq1}
cost_{n_i^m}^{loc} = \gamma k_{n_i^m}{f_{n_i^m}^{n_i^m}}^2C_{n_i^m},
\end{equation}
where $k_{n_i^m}$ denotes the energy consumption coefficient of TD $n_i^m$ determined by the chip architecture, and $\gamma $ denotes the cost of unit energy consumption.

\textit{\textbf{2) Primary ES execution mode:}} In the primary ES execution mode, task $H_{n_i^m}$ will be processed on ES $m$, as shown in Fig. \ref{mode}(a). The delay of task $H_{n_i^m}$ transferring to ES $m$ can be calculated as
\begin{equation}
\label{eq2}
t_{n_i^m,m}^{n_i^m,tran} = \frac{L_{n_i^m}}{R_{n_i^m,m}},
\end{equation}
where $R_{n_i^m,m}=Wlog_2(1+\frac{p_{n_i^m}|h_{n_i^m}|^2}{I_{n_i^m}+N_0})$ denotes the uplink transmission rate of TD $n_i^m$, $W$ is the channel bandwidth, $p_{n_i^m}$ denotes the transmit power of TD $n_i^m$, $h_{n_i^m}$ denotes the channel gain between TD $n_i^m$ and ES $m$, $I_{n_i^m}$ denotes the inter-cell interference to TD $n_i^m$, and $N_0$ is the background noise power. Notably, because BS $m$ and ES $m$ are directly connected via optical fiber, the transmission delay between them is negligible and thus ignored. Considering the maximum tolerated delay of $H_{n_i^m}$, the execution time of this task on ES $m$ cannot exceed $ t_{n_i^m}^{max} - t_{n_i^m,m}^{n_i^m,tran}$, where the delay of the result return is ignored \cite{35}. Therefore, as in the previous derivation process for the local execution mode, the computational resources allocated by ES $m$ for the task $H_{n_i^m}$ are $ f_m^{n_i^m} = \frac{C_{n_i^m}}{t_{n_i^m}^{max} -t_{n_i^m,m}^{n_i^m,tran}}$
. The cost of TD $n_i^m$ in this mode is calculated as
\begin{equation}
\label{eq3}
cost_{n_i^m}^{off}=\gamma p_{n_i^m} t_{n_i^m,m}^{n_i^m,tran}+ \alpha_{n_i^m},
\end{equation}
where $\alpha_{n_i^m}$ is the fee that TD $n_i^m$ needs to pay to the ESP. The cost of ES $m$ is calculated as
\begin{equation}
\label{eq3+}
cost_m^{n_i^m,comp}=\gamma k_m {f_{m}^{n_i^m}}^2C_{n_i^m}.
\end{equation}

\textit{\textbf{3) Neighboring ES execution mode:}} As shown in Fig. \ref{mode}(b), task $H_{n_i^m}$ can be processed on ES $m^{\prime}$, $ \forall m^{\prime} \in \mathcal{M} \backslash \{m\}$ via forwarding from ES $m$. The transmission delay of task $H_{n_i^m}$ from ES $m$ to ES $m^{\prime}$ is calculated as
\begin{equation}
\label{eq4}
t_{m,m^{\prime}}^{n_i^m,tran } = \frac{L_{n_i^m}}{R_{m,m^{\prime}}},
\end{equation}
where $ R_{m,m^{\prime}}$ denotes the transmission rate from ES $m$ to ES $m^{\prime}$. Similar to the previous analysis, the execution time of $H_{n_i^m}$ on ES $m^{\prime}$ is $ t_{n_i^m}^{max} - t_{n_i^m,m }^{n_i^m,tran}-t_{m, m^{\prime}}^{n_i^m,tran }$, which represents the maximum tolerable delay minus the transmission delay of the task offloading from TD $n_i^m$ to ES $m$, and then forwarding it to ES $m^{\prime}$. Therefore, the computational resources allocated by ES $m^{\prime}$ for $H_{n_i^m}$ are $f_{m^{\prime}}^{n_i^m}=\frac{C_{n_i^m}}{t_{n_i^m}^{max} - t_{n_i^m,m }^{n_i^m,tran}-t_{m,m^{\prime}}^{n_i^m,tran }}$.

According to Fig. \ref{mode}(b), TD $n_i^m$ still offloads its task to ES $m$. Therefore, the transmission cost of TD $n_i^m$ in this mode is the same as in the primary ES execution mode. Meanwhile, the fee charged by the ESP to TD $n_i^m$ remains unchanged, regardless of the processing location of its offloaded task. This is because the fee is determined during the initial bargaining game between TD $n_i^m$ and ES $m$, and it does not vary based on the final task scheduling decisions. As a result, TDs only need to offload tasks without concerning the actual processing locations. This ensures fairness among TDs, as they are all charged the same price for the same quality of service, preventing any potential bias that could arise from changes in task processing locations. The transmission cost for ES $m$ is calculated as
\begin{equation}
\label{eq4+}
cost_{m,m^{\prime}}^{n_i^m,tran } = \gamma p_m\frac{L_{n_i^m}}{R_{m,m^{\prime}}},
\end{equation}
where $p_m$ represents the transmission power of ES $m$. The computational cost for ES $m^{\prime}$ is calculated as 
\begin{equation}
\label{eq4++}
cost_{m^{\prime}}^{n_i^m,comp}=\gamma k_{m^{\prime}} {f_{m^{\prime}}^{n_i^m}}^2C_{n_i^m}.
\end{equation}

\textit{\textbf{4) AD execution mode:}} Each ES can delegate some tasks to ADs within its cell and provide rewards to the ADs that assist in handling the tasks, as shown in Fig. \ref{mode}(c). Notably, consistent with \cite{4,5,27}, we assume that ADs can only assist the ES with tasks offloaded by TDs from the same cell. If task $H_{n_i^m}$ is delegated to AD $k_j^m$ for processing, the delay of transferring task $H_{n_i^m}$ from ES $m$ to AD $k_j^m$ is calculated as 
\begin{equation}
\label{eq5}
t_{m,k_j^m}^{n_i^m,tran} = \frac{L_{n_i^m}}{R_{m,k_j^m}},
\end{equation}
where $ R_{m,k_j^m}=Wlog_2(1+\frac{p_m|h_{k_j^m}|^2}{I_{k_j^m}+N_0})$ denotes the transmission rate from ES $m$ to AD $k_j^m$, $h_{k_j^m}$ denotes the channel gain between AD $k_j^m$ and ES $m$, and $I_{k_j^m}$ denotes the inter-cell interference to AD $k_j^m$.

Consistent with the previous analysis, the execution time of task $H_{n_i^m}$ on AD $k_j^m$ is $t_{n_i^m}^{max} - t_{n_i^m,m }^{n_i^m,tran}-t_{m,k_j^m}^{n_i^m,tran}$. The computational resources allocated by AD $k_j^m$ for task $H_{n_i^m}$ are $f_{k_j^m}^{n_i^m}=\frac{C_{n_i^m}}{t_{n_i^m}^{max} - t_{n_i^m,m}^{n_i^m,tran}-t_{m,k_j^m}^{n_i^m,tran}}$. In this mode, the cost of TD $n_i^m$ remains the same, and the cost of ES $m$ is calculated as
\begin{equation}
\label{eq6}
cost_{m, k_j^m}^{n_i^m,tran } = \gamma p_m\frac{L_{n_i^m}}{R_{m, k_j^m}} + \alpha_{k_j^m},
\end{equation}
where $ \alpha_{k_j^m}$ denotes the reward that ES $m$ pays for AD $k_j^m$. The cost of AD $k_j^m$ is calculated as
\begin{equation}
\label{eq6+}
cost_{k_j^m}^{n_i^m,comp}=\gamma k_{k_j^m} {f_{k_j^m}^{n_i^m}}^2C_{n_i^m}.
\end{equation}

\subsection{Utility Model}\label{II-D}
The considered network scenario involves various participants, including TDs, ADs, and the ESP. Next, the utility of each participant is analyzed to facilitate effective collaboration among the three parties.

\textit{\textbf{1) Utility of TDs:}} Based on the cost analysis of TDs in the previous subsection, the utility of TD $n_i^m $ can be expressed as
\begin{equation}
\label{eq7}
u_{n_i^m} = \mathbb{I}(x_{n_i^m}=0)u_{n_i^m}^{loc}+\mathbb{I}(x_{n_i^m}=1) u_{n_i^m}^{off},
\end{equation}
where $\mathbb{I}(\cdot)$ is the indicator function and equals 1 (resp., 0) if the condition is true (resp., false). $u_{n_i^m}^{loc} =  \mathbb{I}(F_{n_i^m} \ge f_{n_i^m}^{n_i^m})(V_{n_i^m}- cost_{n_i^m}^{loc})$ denotes the utility of TD $n_i^m $ to handle its task locally, where $F_{n_i^m}$ denotes the available computational resources of TD $n_i^m$, and $V_{n_i^m}$ denotes the utility that TD $n_i^m $ can obtain when the task $H_{n_i^m}$ is successfully processed.

\textit{\textbf{2) Utility of ADs:}} We use an auction mechanism to incentivize ADs to assist ESs in processing tasks, assuming that the lowest acceptable price reported by AD $k_j^m$ for bidding is $a_{k_j^m}$ $\$$/cycle. When ES $m$ delegates tasks requiring $C_{k_j^m}$ CPU cycles to AD $k_j^m$ and the ESP pays $\alpha_{k_j^m}$ to AD $k_j^m$, the utility of AD $k_j^m$ is calculated as
\begin{equation}
\label{eq7+}
u_{k_j^m}=\alpha_{k_j^m}-a_{k_j^m}C_{k_j^m}.
\end{equation}

\textit{\textbf{3) Utility of the ESP:}} The utility of the ESP is the sum of the utility of all ESs. For tasks processed locally, the ESP can't obtain any utility. For the offloaded task $H_{n_i^m}$, there are three cases to discuss separately.

First, to facilitate the subsequent description, we use $z_{n_i^m}$ to denote the processing location of task $H_{n_i^m}$. Combining the previous definitions of $x_{n_i^m}$ and $y_{n_i^m}$ with the analysis of the four task execution modes, we define the following:
\begin{itemize}
\item{$ z_{n_i^m}= loc $ means $H_{n_i^m}$ is processed locally, \textit{i.e.}, $x_{n_i^m} = 0$ and $y_{n_i^m} = loc$;}
\item{$ z_{n_i^m}= m$ means $H_{n_i^m}$ is processed on ES $m$, \textit{i.e.}, $x_{n_i^m} = 1$ and $y_{n_i^m} = pe$;}
\item{$ z_{n_i^m}= m^{\prime}, \forall m^{\prime} \in \mathcal{M}\backslash \{m\} $ means $H_{n_i^m}$ is processed on ES $m^{\prime}$ , \textit{i.e.}, $x_{n_i^m} = 1$ and $y_{n_i^m} = ne$;}
\item{$z_{n_i^m}= k_j^m, \forall k_j^m \in \mathcal{K}_m$ means $H_{n_i^m}$ is processed on AD $ k_j^m $, \textit{i.e.}, $x_{n_i^m} = 1$ and $y_{n_i^m} = ad$;}
\end{itemize}

According to the above definitions and the cost analysis of ESs in Section \ref{II-C}, if $ z_{n_i^m}= m$, the utility obtained by the ESP is $u_{esp}^{n_i^m,m} = \alpha_{n_i^m}-cost_m^{n_i^m, comp}$. If $ z_{n_i^m}= m^{\prime}, \forall m^{\prime} \in \mathcal{M}\backslash \{m\} $,  the utility obtained by the ESP is $u_{esp}^{n_i^m, m^{\prime}} = \alpha_{n_i^m}- cost_{m,m^{\prime}}^{n_i^m,tran }- cost_{m^{\prime}}^{n_i^m,comp} $. If $ z_{n_i^m}= k_j^m, \forall k_j^m \in \mathcal{K}_m$, the utility obtained by the ESP is $u_{esp}^{n_i^m, k_j^m } = \alpha_{n_i^m}- cost_{m, k_j^m}^{n_i^m,tran } - \beta_{n_i^m}^{k_j^m}$, where $\beta_{n_i^m}^{k_j^m}$ denotes the fee that the ESP pays to AD $k_j^m$ for processing task $H_{n_i^m}$.

Consequently, the utility of the ESP can be calculated as
\begin{equation}
\label{eq7++}
\begin{aligned}
u_{esp} = & \sum_{m \in \mathcal{M}} \sum_{n_i^m \in \mathcal{N}_m} \mathbb{I}(x_{n_i^m} = 1) \bigg\{ \\
& \quad \mathbb{I}(y_{n_i^m} = pe) u_{esp}^{n_i^m, m} + \\
& \quad \mathbb{I}(y_{n_i^m} = ne) \bigg\{ \sum_{m' \in \mathcal{M} \backslash \{m\}} \mathbb{I}(z_{n_i^m} = m') u_{esp}^{n_i^m, m'} \bigg\} + \\
& \quad \mathbb{I}(y_{n_i^m} = ad) \bigg\{ \sum_{k_j^m \in \mathcal{K}_m} \mathbb{I}(z_{n_i^m} = k_j^m) u_{esp}^{n_i^m, k_j^m} \bigg\} \\
& \bigg\}.
\end{aligned}
\end{equation}

\subsection{Problem Formulation}\label{II-E}
ESs, ADs, and TDs form a multilayer edge structure in IoT networks. Our goal is to optimize task allocation and service pricing jointly, improve the utilization of end-device resources, and incentivize efficient collaboration among participants in the network. Therefore, considering the available resources, task demands, and utility of different participants, the optimization problem is formulated to maximize the system utility through optimizing task allocation and service pricing, which is given by
\begin{equation}
\begin{aligned}\label{P0}
\mathcal{P}_1:&\mathop {\emph{max} }\limits_{\boldsymbol{X,Y,Z,A,P}} \; u_{esp}+\sum_{m\in\mathcal{M}}\{\sum_{n_i^m \in \mathcal{N}_m}u_{n_i^m} + \sum_{k_j^m \in \mathcal{K}_m}u_{k_j^m}\} \\
\mbox{s.t.}\;&(a): x_{n_i^m}\in \{0,1\}, \forall m \in \mathcal{M}, \forall n_i^m \in \mathcal{N}_m;  \\
&(b):y_{n_i^m} \in \{loc,pe,ne,ad\}, \forall m \in \mathcal{M}, \forall n_i^m \in \mathcal{N}_m; \\
&(c):z_{n_i^m}= y_{n_i^m} = loc, if \;x_{n_i^m} = 0;\\
&(d):z_{n_i^m} = m, if\; x_{n_i^m} = 1\; and \;y_{n_i^m} = pe;\\
&(e):z_{n_i^m} \in \mathcal{M}\backslash \{m\}, if\; x_{n_i^m} = 1 \;and \;y_{n_i^m} = ne;\\
&(f):z_{n_i^m} \in \mathcal{K}_m, if \;x_{n_i^m} = 1 \;and\; y_{n_i^m} = ad;\\
&(g):\alpha_{n_i^m} \ge 0, \forall m \in \mathcal{M}, \forall n_i^m \in \mathcal{N}_m;\\
&(h):p_{k_j^m} \ge a_{k_j^m}, \forall m \in \mathcal{M}, \forall k_j^m \in \mathcal{K}_m;\\
&(i):\sum_{m\in\mathcal{M}}\sum_{n_i^m\in \mathcal{N}_m} \mathbb{I}({z_{n_i^m}=q}) f_q^{n_i^m} \le F_q, \forall q \in \mathcal{M}\cup \mathcal{K};\\
&(j):u_{esp}\ge 0;\\
&(k):u_{n_i^m} \ge 0, \forall m \in \mathcal{M}, \forall n_i^m \in \mathcal{N}_m;\\
&(l):u_{k_j^m} \ge 0, \forall m \in \mathcal{M}, \forall k_j^m \in \mathcal{K}_m.
\end{aligned}
\end{equation}

The first term of the objective function is the ESP's utility, and the second term is the sum of the utility for all TDs and ADs. $\boldsymbol{X}=\{x_{n_i^m}|m\in\mathcal{M}, n_i^m \in \mathcal{N}_m\}$,  $\boldsymbol{Y}=\{y_{n_i^m}|m\in\mathcal{M}, n_i^m \in \mathcal{N}_m\}$, and $\boldsymbol{Z}=\{z_{n_i^m}|m\in\mathcal{M}, n_i^m \in \mathcal{N}_m\}$ denote the offloading decisions, execution modes, and processing locations of all tasks, respectively. $\boldsymbol{A}=\{\alpha_{n_i^m}|m\in\mathcal{M}, n_i^m \in \mathcal{N}_m\}$ represents the task processing fees that all TDs need to pay to the ESP. $\boldsymbol{P}=\{p_{k_j^m}|m\in\mathcal{M}, k_j^m \in \mathcal{K}_m\}$ represents the reward per CPU cycle that the ESP pays to ADs for assisting in task processing. 

Constraints $(14a)-(14f)$ denote the range for values of the variables related to task allocation. Constraint $(14g)$ ensures that the computation fee paid by each TD to the ESP cannot be negative, and constraint $(14h)$ ensures that the reward per CPU cycle paid by the ESP to any AD assisting in task processing cannot be lower than its minimum acceptable price. Constraint $(14i)$ implies that the total amount of computing resources allocated by each computing node cannot exceed its available resources. Constraints $(14j)-(14l)$ ensure that the utility of the ESP, TDs, and ADs is not negative, i.e., the individual rationality constraint must be satisfied.

\begin{theorem}
The system utility maximization problem $\mathcal{P}_1$ is NP-hard.
\end{theorem}

\newenvironment{proof}{{\noindent\it\quad Proof:}}{\hfill $\square$\par}
\begin{proof}
We consider a specific case of the system utility maximization problem. First, we simplify the network model to a single cell. Then, we assume that the values of $\boldsymbol{X, A, P}$ are given. At this point, the remaining optimization variables are the processing locations of all offloaded tasks. According to the analysis in Section \ref{II-C}, each offloaded task brings different system utility and requires different computational resources when processed on different compute nodes. We can treat each compute node as a backpack with different capacities. Each offloaded task can be treated as an item with different values and weights when placed in different backpacks. Therefore, the problem is transformed into maximizing the total value of the items placed in all backpacks, while ensuring the sum of the weights of the items in each backpack does not exceed its capacity limit. This is known as the multi-knapsack problem, a well-known NP-hard problem \cite{36}. Since this specific case of the problem can be mapped to a multi-knapsack problem, it can be inferred that problem $\mathcal{P}_1$ is also NP-hard.
\end{proof}

According to the above analysis, the problem remains NP-hard even after several simplifications. Therefore, finding a near-optimal solution to the original problem with low computational complexity becomes a key challenge. In the following sections, we will present a multi-level task scheduling framework that is scalable and can be applied to large-scale multi-cell IoT networks.

\begin{figure*}[t]
\centerline{\includegraphics[width=\textwidth]{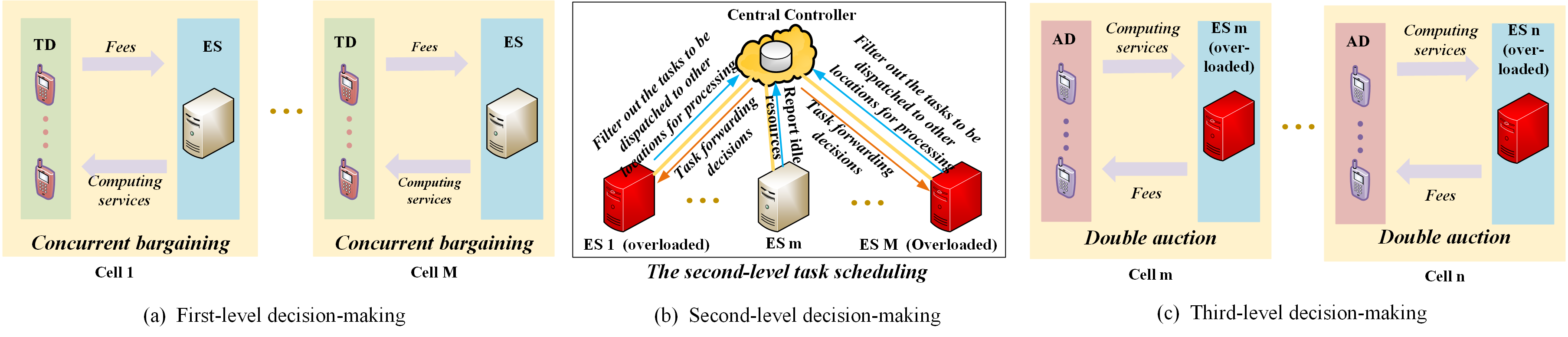}}
    \caption{Logic diagram of the multi-level decision-making framework.}
\label{workflow}
\end{figure*}

\section{Algorithm design}
\subsection{Overview of the Multi-Level Decision-Making Framework}\label{III-A}
This subsection first provides an overall overview of the designed framework. As shown in Fig. \ref{workflow}, the proposed task scheduling framework is divided into three levels. The first level of decision-making uses a bargaining game to determine the initial offloading decision and the computational fee to be paid to the ESP for each TD, as shown in Fig. \ref{workflow}(a). Notably, the first level of decision-making is executed in parallel in each cell. After the first level of decision-making, assuming all the offloaded tasks are using the primary ES execution mode, we can determine whether each ES is overloaded. The overloaded ESs need to screen out tasks that will be dispatched to other locations for processing. As shown in Fig. \ref{workflow}(b), in the second level of decision-making, the overloaded ESs report the screened tasks to the central controller, while the underloaded ESs report their available resources. Subsequently, the central controller decides on the tasks with the neighboring ES execution mode and their specific processing locations. Therefore, if no ES is overloaded after the first level of decision-making, all offloaded tasks will use the primary ES computation mode, and the decision-making process ends. If all ESs are overloaded, the second level of decision-making will be skipped, and the third level of decision-making will be executed directly. As shown in Fig. \ref{workflow}(c), in the third level of decision-making, the ESs that are still overloaded will assign certain tasks to ADs in their cells through a double auction. Finally, tasks that are still not assigned a processing location will be processed locally. Notably, the third level of decision-making is executed in parallel in cells with overloaded ESs. Additionally, as described in Section \ref{II-C}, the fees paid to the ESP by TDs that offload tasks will not change after the first level of decision-making; only the processing location may change.

The following subsections will describe the decision-making process at each level in detail.

\subsection{First Level of Decision-Making}\label{III-B}

As shown in Fig. \ref{workflow}(a), in the first level of decision-making, a parallel bargaining game is employed to determine the initial offloading decision and payment fee for each TD. The first level of decision-making is executed in parallel in each cell, with all TDs and the ES in a cell bargaining simultaneously. This approach not only accelerates the algorithm but also ensures fairness \cite{37}. The parallel game mechanism guarantees fairness among TDs by preventing the outcome of one bargaining round from influencing the next. Moreover, the bargaining game, grounded in its theoretical foundation and the symmetry between both parties, ensures fairness between the TD and ES \cite{+10,+11}.

In a bargaining game, two players are bargaining with each other, either reaching an agreement point in a set $\mathcal{A}$ or reaching a disagreement point $\mathcal{D}$. For the bargaining game between ES $m$ and TD $n_i^m$, the set $\mathcal{A}$ corresponds to the set of feasible computational fees that TD $n_i^m$ pays to ES $m$ for offloading. The disagreement point $\mathcal{D}$ corresponds to the situation where TD $n_i^m$ decides to process the task locally, implying that the bargaining process breaks down. The Nash bargaining solution assigns an outcome to the bargaining game, which can be either an agreement point or a disagreement point. This outcome maximizes the product of the parties' surpluses relative to the disagreement point \cite{24}.

To further reduce the information exchange time introduced by the bargaining game, we solve the Nash bargaining solution for each bargaining game, treating it as the bargaining outcome for both parties. At this point, neither party will have an incentive to change the outcome. Next, we solve the Nash bargaining solution for TD $n_i^m $ and ES $m$. Since the utility of TD $n_i^m $ and ES $m$ at the disagreement point is $u_{n_i^m}^{loc}$ and $0$, respectively, the first level of decision-making problem can be formulated as
\begin{equation}
\begin{aligned}\label{P1}
\mathcal{P}_2:&\mathop {\emph{max} }\limits_{x_{n_i^m},\alpha_{n_i^m}} \;(u_{n_i^m}-u_{n_i^m}^{loc})u_{m}^{n_i^m} \\
\mbox{s.t.}\;&(a):u_{n_i^m}\ge 0,  \\
&(b):u_{m}^{n_i^m} \ge 0, \\
&(c):x_{n_i^m} \in \{0,1\},\\
&(d):\alpha_{n_i^m} \ge 0,
\end{aligned}
\end{equation}
where $u_{m}^{n_i^m}= \mathbb{I}(x_{n_i^m}=1) u_{esp}^{n_i^m,m} $ denotes the gain that ES $m$ can obtain from TD $n_i^m$. Conditions $(15a)$ and $(15b)$ imply that both TD $n_i^m$ and ES $m$ are rational individuals. Conditions $(15c)$ and $(15d)$ specify the range of values for the optimization variables.

In conjunction with Section \ref{II-D}, the problem $\mathcal{P}_2$ can be reformulated as
\begin{equation}
\begin{aligned}\label{P2}
\mathcal{P}_3:&\mathop {\emph{max} }\limits_{x_{n_i^m},\alpha_{n_i^m}} ((1-x_{n_i^m})u_{n_i^m}^{loc}+x_{n_i^m} u_{n_i^m}^{off} - u_{n_i^m}^{loc})x_{n_i^m}u_{esp}^{n_i^m,m} \\
\mbox{s.t.}\;&(a):(15a)-(15d).
\end{aligned}
\end{equation}

When  $ x_{n_i^m} = 0$, it can be deduced that $ \alpha_{n_i^m} = 0$, $u_{n_i^m}\ge 0$, $u_{m}^{n_i^m} = 0$, and the value of the objective function is zero. When $ x_{n_i^m} = 1$, the problem $\mathcal{P}_3$ transforms to
\begin{equation}
\begin{aligned}\label{P3}
\mathcal{P}_4:&\mathop {\emph{max} }\limits_{\alpha_{n_i^m}} \;(u_{n_i^m}^{off}- u_{n_i^m}^{loc})u_{esp}^{n_i^m,m} \\
\mbox{s.t.}\;&(a):u_{n_i^m}^{off} \ge 0,  \\
&(b):u_{esp}^{n_i^m,m} \ge 0, \\
&(c):(15d).
\end{aligned}
\end{equation}

Therefore, $ x_{n_i^m} = 1$ is a Nash solution if and only if $ u_{n_i^m}^{off}- u_{n_i^m}^{loc} > 0 $ and $ u_{esp}^{n_i^m,m}>0 $. Next, we need to determine the optimal value of $\alpha_{n_i^m}$ when $ x_{n_i^m} = 1$ is a Nash solution. First, we expand the previous two inequalities to obtain
\begin{equation}
\label{eq8}
V_{n_i^m}- \gamma p_{n_i^m} t_{n_i^m,m}^{n_i^m,tran}- \alpha_{n_i^m} - u_{n_i^m}^{loc} > 0,
\end{equation}
and
\begin{equation}
\label{eq9}
\alpha_{n_i^m}-\gamma k_m {f_{m}^{n_i^m}}^2C_{n_i^m} > 0.
\end{equation}
Then, we can determine the range of values for $\alpha_{n_i^m}$ when $ x_{n_i^m} = 1$ is a Nash solution:
\begin{equation}
\label{eq10}
\alpha_{n_i^m}^{min}<\alpha_{n_i^m}< \alpha_{n_i^m}^{max},
\end{equation}
where $\alpha_{n_i^m}^{min}=\gamma k_{m}{f_{m}^{n_i^m}}^2C_{n_i^m}$ and $ \alpha_{n_i^m}^{max}=V_{n_i^m}- \gamma p_{n_i^m} t_{n_i^m,m}^{n_i^m,tran}-u_{n_i^m}^{loc}$.

Therefore, $x_{n_i^m}^1 = 0$ and $\alpha_{n_i^m}^1=0 $ constitute the Nash bargaining solution between TD $n_i^m$ and ES $m$ when $\alpha_{n_i^m}^{min} \ge \alpha_{n_i^m}^{max}$. Additionally, the values of the optimization variables associated with the task $H_{n_i^m}$ are determined as $x_{n_i^m}^*=0, y_{n_i^m}^* =loc, z_{n_i^m}^* = loc, \alpha_{n_i^m}^* = 0$. When $\alpha_{n_i^m}^{min} < \alpha_{n_i^m}^{max}$, $x_{n^m_i}^1 = 1$ is a Nash solution. At this point, we can expand the objective function of $\mathcal{P}_4$ as follows:
\begin{equation}
\label{eq11}
- \alpha_{n_i^m}^2 +B\alpha_{n_i^m} +C,
\end{equation}
where $ B = V_{n_i^m} - \gamma p_{n_i^m} t_{n_i^m,m}^{n_i^m,tran} - u_{n_i^m}^{loc}+ \gamma k_m {f_m^{n_i^m}}^2 C_{n_i^m}$ and $ C = - V_{n_i^m} \gamma k_m {f_m^{n_i^m}}^2 C_{n_i^m} + \gamma^2 p_{n_i^m} t_{n_i^m,m}^{n_i^m,tran} k_m {f_m^{n_i^m}}^2 C_{n_i^m} + \gamma u_{n_i^m}^{loc} k_m {f_m^{n_i^m}}^2 C_{n_i^m}$. Therefore, equation (\ref{eq11}) is a typical univariate quadratic concave function, meaning it has a unique maximum point. Combined with the range of values for $\alpha_{n_i^m}$, we can ascertain the optimal solution for $\alpha_{n_i^m}$ as follows:
\begin{equation}
\label{eq12}
\alpha_{n_i^m}^1 = \begin{cases} \frac{1}{2}B & \text{if }\; \alpha_{n_i^m}^{min}<\frac{1}{2}B<\alpha_{n_i^m}^{max}, \\ \alpha_{n_i^m}^{min} & \text{if } \; \alpha_{n_i^m}^{min} \ge \frac{1}{2}B,\\ \alpha_{n_i^m}^{max} & \text{if } \;\alpha_{n_i^m}^{max} \le \frac{1}{2}B.\end{cases}
\end{equation}

At the first level of decision-making, we obtain the initial offloading decision and payment fee for each TD. As described in Section \ref{III-A}, we also need to determine whether each ES is overloaded. For the overloaded ESs, we need to filter out the tasks to be dispatched to other locations for processing.

To determine whether each ES is overloaded, we assume that all tasks decided to be offloaded in the first level of decision-making will use the primary ES execution mode. Then, the amount of computational resources that ES $m, \forall m \in \mathcal{M}$ needs to provide can be expressed as
\begin{equation}
\label{eq13}
f_{m}^1 = \sum_{n_i^m \in \mathcal{N}_m} x_{n_i^m}^{*}f_{m}^{n_i^m}.
\end{equation}
If $f_{m}^1 > F_m$, it indicates that ES $m$ is overloaded. Otherwise, ES $m$ is underloaded. For the underloaded ESs, all offloaded tasks in their cells will employ the primary ES execution mode. Therefore, the optimization variables for these tasks are determined as $x_{n_i^m}^* = 1, y_{n_i^m}^*=pe, z_{n_i^m}^*=m, \alpha_{n_i^m}^* = \alpha_{n_i^m}^1,  \forall n_i^m \in \{n_i^m| f_m^1 \ge F_m, x_{n_i^m}=1\}$.

For the overloaded ES $m$, we need to filter out tasks to be dispatched to other locations for processing. For this purpose, we design a priority-based task filtering algorithm, where the priority function is defined as
\begin{equation}
\label{eq14}
O_{n_i^m} = \frac{u_m^{n_i^m}}{f_m^{n_i^m}}.
\end{equation}
A larger value of the priority function indicates that ES $m$ gains more utility per unit of computational resource by processing the task locally. Hence, the task is given higher priority to execute on ES $m$. The designed task filtering algorithm is presented in Algorithm \ref{algo1}. 

In Algorithm \ref{algo1}, $\mathcal{I}_m$ denotes the set of indexes for tasks to be offloaded in the first level of decision-making. $F_m$ denotes the total amount of computational resources for ES $m$. $\mathcal{I}_m ^{loc}$ denotes the set of indexes for tasks to be processed at ES $m$, and $\mathcal{I}_m ^{off}$ denotes the set of indexes for tasks to be dispatched to other locations for processing. Furthermore, according to the description in Sec \ref{III-A}, Algorithm 1 is executed in parallel across all overloaded ESs. All tasks in $\mathcal{I}_m ^{loc}$ will use the primary ES execution mode, with the optimization variables for these tasks determined as $x_{n_i^m}^* = 1, y_{n_i^m}^*=pe, z_{n_i^m}^*=m, \alpha_{n_i^m}^* = \alpha_{n_i^m}^1,\forall i \in \mathcal{I}_m ^{loc}$.

\IncMargin{1em}
\begin{algorithm} \SetKwData{Left}{left}\SetKwData{This}{this}\SetKwData{Up}{up} \SetKwFunction{Union}{Union}\SetKwFunction{FindCompress}{FindCompress} \SetKwInOut{Input}{input}\SetKwInOut{Output}{output}
	
	\Input{$\mathcal{I}_m$ and $F_m$.} 
	\Output{$\mathcal{I}_m ^{loc}$ and $\mathcal{I}_m ^{off}$.}
	\BlankLine 

    Initialize $f_m^{left} = F_m, \mathcal{I}_m^{loc} = \phi, \mathcal{I}_m^{off}=\phi$\;
    Calculate the priority function value of all tasks in $\mathcal{I}_m$ according to (\ref{eq14}). Then, rank the task indexes in descending order based on these values to obtain  $\mathcal{I}_m^{\prime}$\;
	\For{$i$ in $\mathcal{I}_m ^{\prime}$}{ 
        \If{$f_m^{n_i^m} \le f_m^{left}$}{
       $f_m^{left} = f_m^{left} - f_m^{n_i^m}$\; 
       $\mathcal{I}_m ^{loc} = \mathcal{I}_m ^{loc} \cup \{i\}$\;
    }
    \Else{ 
            $\mathcal{I}_m ^{off} =  \mathcal{I}_m ^{off} \cup \{i\}$\;
        }      
    } 
    $\mathbf{return}$ $\mathcal{I}_m ^{loc}$, $\mathcal{I}_m ^{off}$
    
    \caption{Priority-Based Task Filtering Algorithm (For overloaded ES $m \in \mathcal{M}$)}\label{algo1} 
    \end{algorithm}
\DecMargin{1em} 

\subsection{Second Level of Decision-Making}\label{III-c}

As described in Section \ref{III-A}, the second level of decision-making identifies the tasks that will use the neighboring ES execution mode and their specific processing locations. At this level, we categorize all ESs into two sets: $\mathcal{M}^{adequate}$ (the set of underloaded ESs) and $\mathcal{M}^{inadequate}$ (the set of overloaded ESs) based on the first level of decision-making. Next, we implement a two-tier prioritization-based decision-making approach. Specifically, in the outer layer, we prioritize the ESs in the set $\mathcal{M}^{inadequate}$. Subsequently, the inner layer decisions are made sequentially based on these established priorities. The priority function for ES $m_2 \in \mathcal{M}^{inadequate}$ is defined as
\begin{equation}
\label{eq15}
Q_{m_2} = f_{m_2}^1 – F_{m_2}.
\end{equation}
The larger the value of $Q_{m_2}$, the more severely ES $m_2$ is overloaded, resulting in a higher priority.

In the inner layer decision for ES $m_2$, we use a greedy algorithm to determine if each task in $I_{m_2}^{off}$ should adopt the neighboring ES execution mode. When a task is determined to employ the  neighboring ES execution mode, its specific processing location also needs to be determined. The tasks that are not assigned processing locations in the second level of decision-making will be addressed in the third level. The process for the second level of decision-making is detailed in Algorithm \ref{algo2}.

\IncMargin{1em}
\begin{algorithm} \SetKwData{Left}{left}\SetKwData{This}{this}\SetKwData{Up}{up} \SetKwFunction{Union}{Union}\SetKwFunction{FindCompress}{FindCompress} \SetKwInOut{Input}{input}\SetKwInOut{Output}{output}
	
	\Input{$f_{m_1}^{left}, \forall m_1 \in \mathcal{M}^{adequate}$, and $\mathcal{I}_{m_2} ^{off}, \forall m_2 \in \mathcal{M}^{inadequate}$.} 
	\Output{$\mathcal{I}_{m_2}^{ne}, \forall m_2 \in \mathcal{M}^{inadequate} $, and $z_{n_i^{m_2}}, \forall i \in \mathcal{I}_{m_2}^{ne}, \forall m_2 \in \mathcal{M}^{inadequate}$.}
	\BlankLine 

    Initialize $\mathcal{I}_{m_2}^{ne} = \phi, \forall m_2 \in \mathcal{M}^{inadequate} $\;
    Calculate the priority of the ESs in $M^{inadequate}$ according to (\ref{eq15}), then sort them in descending order to obtain ${\mathcal{M}^{inadequate }}^{\prime}$\;
	\For{${m_2}^{\prime}$ in ${\mathcal{M}^{inadequate}}^{\prime}$}{
        Sort $m_1 \in \mathcal{M}^{adequate}$ in descending order of $R_{{m_2}^{\prime}, {m_1}}$ to yield ${\mathcal{M}^{adequate}}^{\prime}$\;
        \For{$i$ in $\mathcal{I}_{{m_2}^{\prime}} ^{off}$}{
            \For{${m_1}^{\prime}$ in ${\mathcal{M}^{adequate}}^{\prime}$}{
                \If{$f_{{m_1}^{\prime}}^{n_i^{{m_2}^{\prime}}} \le f_{{m_1}^{\prime}}^{left} $ and $u_{{m_1}^{\prime}}^{n_i^{{m_2}^{\prime}}}>0$}{
                    $\mathcal{I}_{{m_2}^{\prime}}^{ne} = \mathcal{I}_{{m_2}^{\prime}}^{ne} \cup \{n_i^{{m_2}^{\prime}}\}$, $z_{n_i^{{m_2}^{\prime}}} = {m_1}^{\prime}$, $ f_{{m_1}^{\prime}}^{left}  = f_{{m_1}^{\prime}}^{left}  -f_{{m_1}^{\prime}}^{n_i^{{m_2}^{\prime}}} $\;
                    break\;
                }
            }
        }
    }
   
    $\mathbf{return}$ $\mathcal{I}_{m_2}^{ne}, \forall m_2 \in \mathcal{M}^{inadequate} $, and $z_{n_i^{m_2}}, \forall i \in \mathcal{I}_{m_2}^{ne}, \forall m_2 \in \mathcal{M}^{inadequate}$
    \caption{Priority-Based Task Scheduling Algorithm Among ESs}\label{algo2} 
    \end{algorithm}
\DecMargin{1em} 

In Algorithm \ref{algo2}, $f_{m_1}^{left}$ represents the amount of idle resources for ES $m_1 \in \mathcal{M}^{adequate}$ after the first level of decision-making. $I_{m_2}^{ne}$ denotes the set of tasks that are decided to use the neighboring ES execution mode after the second level of decision-making. $z_{n_i^{m_2}}$ represents the specific processing location of task $H_{n_i^{m_2}}$, where $i \in \mathcal{I}_{m_2}^{ne}$ and $m_2 \in \mathcal{M}^{inadequate}$. After the second level of decision-making, all optimization variables for tasks employing the neighboring ES execution mode are determined as $x_{n_i^{m_2}}^*=1,y_{n_i^{m_2}}^*=ne,z_{n_i^{m_2}}^*=z_{n_i^{m_2}},\alpha_{n_i^{m_2}}^*=\alpha_{n_i^{m_2}}^1, \forall i \in \mathcal{I}_{m_2}^{off},\forall m_2 \in \mathcal{M}^{inadequate}$.

\subsection{Third Level of Decision-Making}\label{III-d}
If there are tasks that remain without a specific processing location after the second level of decision-making, or if all ESs are overloaded after the first level of decision-making, the corresponding cells will initiate the third level of decision-making. In this case, the third level of decision-making is executed in parallel across all relevant cells. The ESs in these cells will delegate certain tasks without processing locations to the ADs for processing, providing a certain reward, as shown in Fig. 3(c). The interaction between the ES and ADs in each cell can be modeled as a double auction. In the cell triggering the third level of decision-making, ADs act as sellers, the ES as the buyer, and the BS as the third-party auctioneer. For clarity, we will use cell $m_3$ as an example to discuss the specific process of the third-level decision.

At the beginning of the double auction, AD $k_j^{m_3}, \forall k_j^{m_3} \in \mathcal{K}_{m_3}$ reports its available computational resources $F_{k_j^{m_3}}$ to BS $m_3$ and bids with the lowest acceptable price $a_{k_j^{m_3}}$ per CPU cycle. Assume the set of task indexes delegated by ES $m_3$ in the third level of decision-making is $\mathcal{I}_{m_3}^{off}$. ES $m_3$ also reports to BS $m_3$ the payment fee for each task in $\mathcal{I}_{m_3}^{off}$ determined in the first level of decision-making. Considering the transmission overhead when ES $m_3$ forwards task $H_{n_i^{m_3}}$ to AD $k_j^{m_3}$, BS $m_3$ can calculate the maximum acceptable payment fee per CPU cycle for ES $m_3$ to delegate task $H_{n_i^{m_3}}$ to AD $k_j^{m_3}$, which is formulated as
\begin{equation}
\label{eq16}
b_{n_i^{m_3}}^{k_{j}^{m_3}} = \frac{\alpha_{n_i^{m_3}}^1 - cost_{{m_3}, k_j^{m_3}}^{n_i^{m_3},tran}}{C_{n_i^{m_3}}}. 
\end{equation}
Subsequently, BS $m_3$ needs to determine the set of tasks using the AD execution mode, their specific processing locations, and the reward ES $m_3$ needs to pay to each AD.

According to \cite{38} and \cite{39}, the desired properties of the double auction design are:
\begin{itemize}
\item{\textit{Individual rationality:} All auction matching outcomes do not compromise participant interests. If task $ H_{n_i^{m_3}}$ is delegated to AD $k_j^{m_3}$ in the third level of decision-making and ES $m_3$ pays a fee of $\beta_{n_i^{m_3}}^{k_j^{m_3}}$ to AD $k_j^{m_3}$ for this task, it should satisfy $ b_{n_i^{m_3}}^{k_j^{m_3}}  \ge \beta_{n_i^{m_3}}^{k_j^{m_3}}/ C_{n_i^{m_3}} $ and $\beta_{n_i^{m_3}}^{k_j^{m_3}}/ C_{n_i^{m_3}} \ge a_{k_j^{m_3}}$.}
\item{\textit{Honesty:} Neither buyers nor sellers have an incentive to alter their bids or offers. All traders submit bids or offers based on their true valuation of the resource. Untruthful offers or bids do not generate additional revenue.}
\item{\textit{Budget Balancing:} After the transaction, the total expenses of all buyers balance with the total income of all sellers, ensuring the auctioneer does not lose money.}
\item{\textit{System Efficiency:} This involves maximizing the number of successful transactions at the end of the auction.}
\end{itemize}

To achieve these goals, we design a double auction matching mechanism based on bids, offers, and resource constraints, presented in Algorithm \ref{algo3}. 

\IncMargin{1em}
\begin{algorithm} \SetKwData{Left}{left}\SetKwData{This}{this}\SetKwData{Up}{up} \SetKwFunction{Union}{Union}\SetKwFunction{FindCompress}{FindCompress} \SetKwInOut{Input}{input}\SetKwInOut{Output}{output}
	
	\Input{$\mathcal{I}_{m_3}^{off}$, $b_{n_i^{m_3}}^{k_j^{m_3}}$,  $a_{k_j^{m_3}}$, and $F_{k_j^{m_3}}$, $\forall i \in \mathcal{I}_{m_3}^{off}$, $\forall k_j^{m_3} \in \mathcal{K}_{m_3} $.} 
	\Output{$\mathcal{G}_{m_3}$, and $\beta_{n_i^{m_3}}^{k_j^{m_3}}$, $\forall i \in \mathcal{I}_{m_3}^{off}$, $\forall k_j^{m_3} \in \mathcal{K}_{m_3} $.}
	\BlankLine 

    Initialize $\mathcal{G}_{m_3} = \phi$, $f_{ k_j^{m_3}}^{left}= F_{k_j^{m_3}}$, $\beta_{n_i^{m_3}}^{k_j^{m_3}}=0$, $\forall i \in \mathcal{I}_{m_3}^{off}$, $\forall k_j^{m_3} \in \mathcal{K}_{m_3} $\;
    Arrange the ADs in ascending order of $a_{k_j^{m_3}}$, yielding $\mathcal{K}_{m_3}^{\prime}$\;
    \For{$i$ in $\mathcal{I}_{m_3}^{off}$}{
        \For{$j = 0, 1, ..., |\mathcal{K}_{m_3}^{\prime}| - 2$}{
            \If{$b_{n_i^{m_3}}^{\mathcal{K}_{m_3}^{\prime}[j]} > a_{\mathcal{K}_{m_3}^{\prime}[j+1]}$ and $f_{\mathcal{K}_{m_3}^{\prime}[j]}^{ n_i^{m_3} } \le f_{ \mathcal{K}_{m_3}^{\prime}[j]}^{left}$}{
                $\mathcal{G}_{m_3} = \mathcal{G}_{m_3} \cup \{(n_i^{m_3}, \mathcal{K}_{m_3}^{\prime}[j])\}$, $f_{ \mathcal{K}_{m_3}^{\prime}[j]}^{left} = f_{ \mathcal{K}_{m_3}^{\prime}[j]}^{left} - f_{\mathcal{K}_{m_3}^{\prime}[j]}^{ n_i^{m_3} }$, $\beta_{n_i^{m_3}}^{\mathcal{K}_{m_3}^{\prime}[j]} = a_{\mathcal{K}_{m_3}^{\prime}[j+1]}* C_{n_i^{m_3}} $\;
                break\;
            }
        }
    }
	
    $\mathbf{return}$ $\mathcal{G}_{m_3}$, $\beta_{n_i^{m_3}}^{k_j^{m_3}}$, $\forall i \in \mathcal{I}_{m_3}^{off}$, $\forall k_j^{m_3} \in \mathcal{K}_{m_3} $
    
    \caption{Double Auction-Based Matching and Pricing Algorithm (Use cell $m_3$ as an example)}\label{algo3} 
    \end{algorithm}
\DecMargin{1em} 

In Algorithm \ref{algo3}, $\mathcal{I}_{m_3}^{off}$ represents the set of task indexes in cell $m_3$ participating in the third level of decision-making, and $\mathcal{G}_{m_3}$ denotes the set of matched pairs of tasks and ADs in cell $m_3$. Similar to the analysis in \cite{38}, the designed algorithm satisfies the economic principles of individual rationality, honesty, and budget balancing, which are not expanded here. The third level of decision-making identifies tasks that employ the AD execution mode and their specific execution locations. Additionally, the reward paid by the ESP to AD $k_j^{m_3}$ is calculated as
\begin{equation}
\label{eq16+}
\alpha_{k_j^{m_3}} =\sum_{ i \in \mathcal{I}_{m_3}^{off}}\beta_{n_i^{m_3}}^{k_j^{m_3}}. 
\end{equation}

After the third level of decision-making, the optimization variables for tasks determined to use the AD execution mode are identified. Finally, tasks with undetermined processing locations are designated to employ the local execution mode. At this point, the values of all optimization variables are specified.

\subsection{Multi-Level Decision-Making Framework}\label{III-e}
In summary, the first level of decision-making identifies tasks using the primary ES execution mode, the second level identifies tasks using the neighboring ES execution mode, the third level identifies tasks using the AD execution mode, and all remaining tasks use the local execution mode. The overall framework is presented in Algorithm \ref{algo4}.

\IncMargin{1em}
\begin{algorithm}[t] \SetKwData{Left}{left}\SetKwData{This}{this}\SetKwData{Up}{up} \SetKwFunction{Union}{Union}\SetKwFunction{FindCompress}{FindCompress} \SetKwInOut{Input}{input}\SetKwInOut{Output}{output}
	\CommentSty{// The process in lines 2-7 is executed in parallel across all cells. The following process is exemplified by cell $m$}

    A parallel bargaining game is played between TD $n_i^m, \forall n_i^m \in \mathcal{N}_m$ and ES $m$. The initial offloading decision $x_{n_i^m}$ and the payment fee $\alpha_{n_i^m}$ of TD $n_i^m$ are determined according to the formulas in Section \ref{III-B}\;
    Calculate $f_m^1$ according to (\ref{eq13})\;
    \If{$f_m^1 > F_m$}{
        Filter the tasks to be dispatched to other locations for processing according to Algorithm \ref{algo1}, and identify the tasks employing the primary ES execution mode\;
    }
    \Else{
        All tasks determined to be offloaded in the first level of decision-making employ the primary ES execution mode and calculate the remaining idle resources of ES $m$\;
    }
    \CommentSty{// The process in lines 9-16 is performed by the central controller}

    \If{all ESs are underloaded}{
        Algorithm ends\;
    }
    \If{all ESs are overloaded}{
        $\mathbf{goto}$ line 19\;
    }
    \Else{
        Execute Algorithm \ref{algo2} to determine the tasks employing the neighboring ES mode and their specific processing locations\;
    }
    \If{All tasks are assigned processing locations}{
        Algorithm ends\;
    }
    \Else{
        \CommentSty{// Line 19 is executed in parallel for cells where tasks with unallocated processing locations exist. The following process is exemplified by cell $m$}

        In cell $m$, execute Algorithm \ref{algo3} to determine (1) the tasks employing the AD execution mode, (2) their specific processing locations, and (3) the fees the ESP needs to pay to the ADs\;
    }
    Tasks without assigned processing locations are determined to employ the local execution mode.
    \caption{Multi-Level Decision-Making Algorithm}\label{algo4} 
    \end{algorithm}
\DecMargin{1em} 

Notably, despite the complexity of the proposed multi-level decision-making framework, only the relevant components need to be implemented on the BSs, ESs and central controller during practical deployment. The components deployed on the BSs handle the first and third-level decisions. The components deployed on the ESs need to provide necessary information for the second-level decision to the central controller component.

At the first level of decision-making, the BS in each cell collects information from all TDs and the ES within the same cell, then calculates the Nash bargaining solutions using the formulas provided in Section\ref{III-B} (line 2). BSs then send these solutions to the corresponding ESs. Afterward, each ES evaluates its load based on these Nash negotiation solutions with all TDs, determines the tasks employing the primary ES execution mode, and sends details of tasks ready for processing elsewhere or information about idle resources to the central controller (lines 3-7).

Based on reports from all ESs, the central controller then decides whether to conclude the decision-making process, skip the second level of decision-making, or proceed with the second level of decision-making (lines 9-13).

If the central controller opts for the second level of decision-making, it determines the tasks employing the neighboring ES mode and their specific processing locations, using task details and idle resources provided by each ES. These decisions are then communicated to the respective ESs (line 14).

Next, the central controller checks if processing locations have been assigned for all tasks. If not, it instructs the ESs with tasks that have unallocated processing locations to start the third level of decision-making, while notifying the others to skip this step. If all tasks have been assigned, the process ends, and all ESs are informed to skip the third level (lines 15-17).

Each ES involved in the third-level of decision-making determines the tasks employing the AD execution mode and their specific processing locations through a double auction organized by the corresponding BS, acting as the auctioneer. Then, tasks without assigned processing locations are determined to employ the local execution mode (lines 19-20). Finally, all ESs transmit task processing decisions to the respective TDs in their cells. Relevant fees will be collected from the respective TDs and paid to the corresponding ADs upon completion of task processing, based on the decision results.

\section{Theoretical Analysis}
In this section, we theoretically analyze the performance of the proposed scheme. First, we analyze the complexity of the proposed scheme. Next, we discuss the equivalence between the decomposed subproblems and the original problem, followed by an analysis of the scheme's optimality.
\subsection{Computational Complexity}\label{IV-A}
The proposed framework employs hierarchical decision-making and parallel execution to reduce computational complexity. The first level of decision-making is executed in parallel for individual cells. Its computational complexity is $\mathcal{O}(1)$ since the Nash bargaining solution is obtained directly via formula calculation. For task screening of overloaded ESs, Algorithm \ref{algo1} has a computational complexity of $\mathcal{O}(NlogN)$, where $N = max(N_1,…,N_M)$. Algorithm \ref{algo2} has a computational complexity of $\mathcal{O}(MlogM+MNlogN)$. Algorithm \ref{algo3} is executed in parallel for certain cells; its computational complexity is $\mathcal{O}(NKlogK)$, where $K = max(K_1,…,K_M)$. Combining the above analyses, the overall computational complexity of the proposed framework is $\mathcal{O}( MlogM+MNlogN+NKlogK)$. 

\subsection{Equivalence Analysis}\label{IV-B}
The system utility can be expressed as the sum of the utilities obtained by TDs, ADs, and the ESP for each task. As stated in Section \ref{II-C}, the fee paid to the ESP when a TD offloads its task is independent of the processing location. Therefore, when calculating the utility for each TD, we need to consider only two scenarios: offloading the task or processing it locally. As mentioned in Section \ref{II-D}, the ESP and ADs receive varying utility for each offloaded task depending on the execution mode. Therefore, the utility of the ESP and ADs needs to be calculated separately for each execution mode. Based on (\ref{eq7}) and (\ref{eq7++}), the objective function of Problem $\mathcal{P}_1$ can be reformulated as follows:
\begin{equation}
\label{eq7+++}
\begin{aligned}& \sum_{m \in \mathcal{M}} \sum_{n_i^m \in \mathcal{N}_m} \mathbb{I}(x_{n_i^m} = 1) \bigg\{ \\& \quad \mathbb{I}(y_{n_i^m} = pe) u_{esp}^{n_i^m, m} + \\& \quad \mathbb{I}(y_{n_i^m} = ne) \bigg\{ \sum_{m' \in \mathcal{M} \backslash \{m\}} \mathbb{I}(z_{n_i^m} = m') u_{esp}^{n_i^m, m'} \bigg\} + \\& \quad \mathbb{I}(y_{n_i^m} = ad) \bigg\{ \sum_{k_j^m \in \mathcal{K}_m} \mathbb{I}(z_{n_i^m} = k_j^m) (u_{esp}^{n_i^m, k_j^m} +u_{k_j^m}^{n_i^m} )\bigg\} + \\&\quad u_{n_i^m}^{off} \\& \bigg\} +\sum_{m \in \mathcal{M}} \sum_{n_i^m \in \mathcal{N}_m} \mathbb{I}(x_{n_i^m} = 0)u_{n_i^m}^{loc}, \end{aligned}
\end{equation}
where $u_{k_j^m}^{n_i^m}$ represents the utility that AD $k_j^m$ gains by assisting ES $m$ in processing task $H_{n_i^m}$. In our proposed multi-level decision-making framework (Section \ref{III-e}), the first level of decision-making identifies tasks using the primary ES execution mode, their processing locations, and the fee for offloading each task. The second level determines tasks that use the neighboring ES execution mode and their processing locations. The third level identifies tasks processed using the AD execution mode, their processing locations, and the reward per CPU cycle paid by the ESP to ADs for assistance. Lastly, tasks processed locally are determined. Therefore, the subproblems solved at each level correspond to subterms in (\ref{eq7+++}), and the final solution includes the optimization variables of the original problem. Moreover, the solutions of all subproblems meet the constraints of problem $\mathcal{P}_1$. Based on the above analysis, the decomposed subproblems are jointly equivalent to the original problem.
\subsection{Optimality Analysis}\label{IV-c}
Based on the previous subsection's analysis, the decomposed subproblems in our multi-level decision-making framework are jointly equivalent to the original problem.  Thus, to analyze optimality, we focus on each level of decision-making individually. In the first level, a parallel bargaining game is employed to determine the initial offloading decision and payment fee for each TD. Decisions are made by solving the Nash bargaining solution, whose optimality is supported by four core Nash axioms: pareto efficiency, symmetry, invariance to affine transformations, and independence of irrelevant alternatives. More details can be found in \cite{+12}. In the second level, a priority-based scheduling algorithm is applied, where priority is determined by the utility per unit of resource. This is essentially a greedy algorithm. Although the greedy algorithm may not achieve a global optimum, it provides an approximate optimal solution with strong performance guarantees. The third level employs a double auction mechanism. Similar to the analysis in \cite{38}, we can show that the algorithm is budget-balanced, individually profitable, system-efficient, and truthful. From the above analyses, it is evident that the proposed framework can achieve an approximate optimal solution to the original problem with strong performance guarantees. We prioritize efficiency in practice, accepting some deviation from the global optimum as it remains within acceptable limits.
\begin{table}[h]
\caption{Main Simulation Parameters}
\label{tab2}
\resizebox{\columnwidth}{!}{%
\begin{tabular}{l|l}
\hline
Parameters                                                                 & Value                    \\ \hline
$M$                                                                        & 5                        \\
$N_m, \forall m \in \mathcal{M}$                                           & {[}200, 800{]}            \\
$K_m, \forall m \in \mathcal{M}$                                            & {[}20,  40{]}             \\
$N_0$                                                                      & -100 dBm                 \\
$k_{n_i^m}, \forall n_i^m \in \mathcal{N}_m, \forall m \in \mathcal{M}$       & $ 5 \times10 ^ {-27} $        \\
$W $                                                                       & 20 MHz                   \\
$F_{n_i^m}, \forall n_i^m \in \mathcal{N}_m, \forall m \in \mathcal{M} $      & {[}0.05, 1{]} GHz         \\
$F_m, \forall m \in \mathcal{M} $                                           & 10 GHz                   \\
$L_{n_i^m}, \forall n_i^m \in \mathcal{N}_m, \forall m \in \mathcal{M} $      & {[}200, 500{]} Kbit       \\
$C_{n_i^m}, \forall n_i^m \in \mathcal{N}_m, \forall m \in \mathcal{M} $      & {[}50, 500{]} Mega Cycles \\
$t_{n_i^m}^{max}, \forall n_i^m \in \mathcal{N}_m, \forall m \in \mathcal{M}$ & {[}0.05, 2{]} s          \\
$\gamma$                                                                   & 1 $\$$                   \\
$V_{n_i^m}, \forall n_i^m \in \mathcal{N}_m, \forall m \in \mathcal{M} $      & {[}5, 10{]} $\$$         \\ \hline
\end{tabular}%
}
\end{table}

\begin{figure*}[b]
\centering
\subfloat[]{\includegraphics[width=0.33\textwidth]{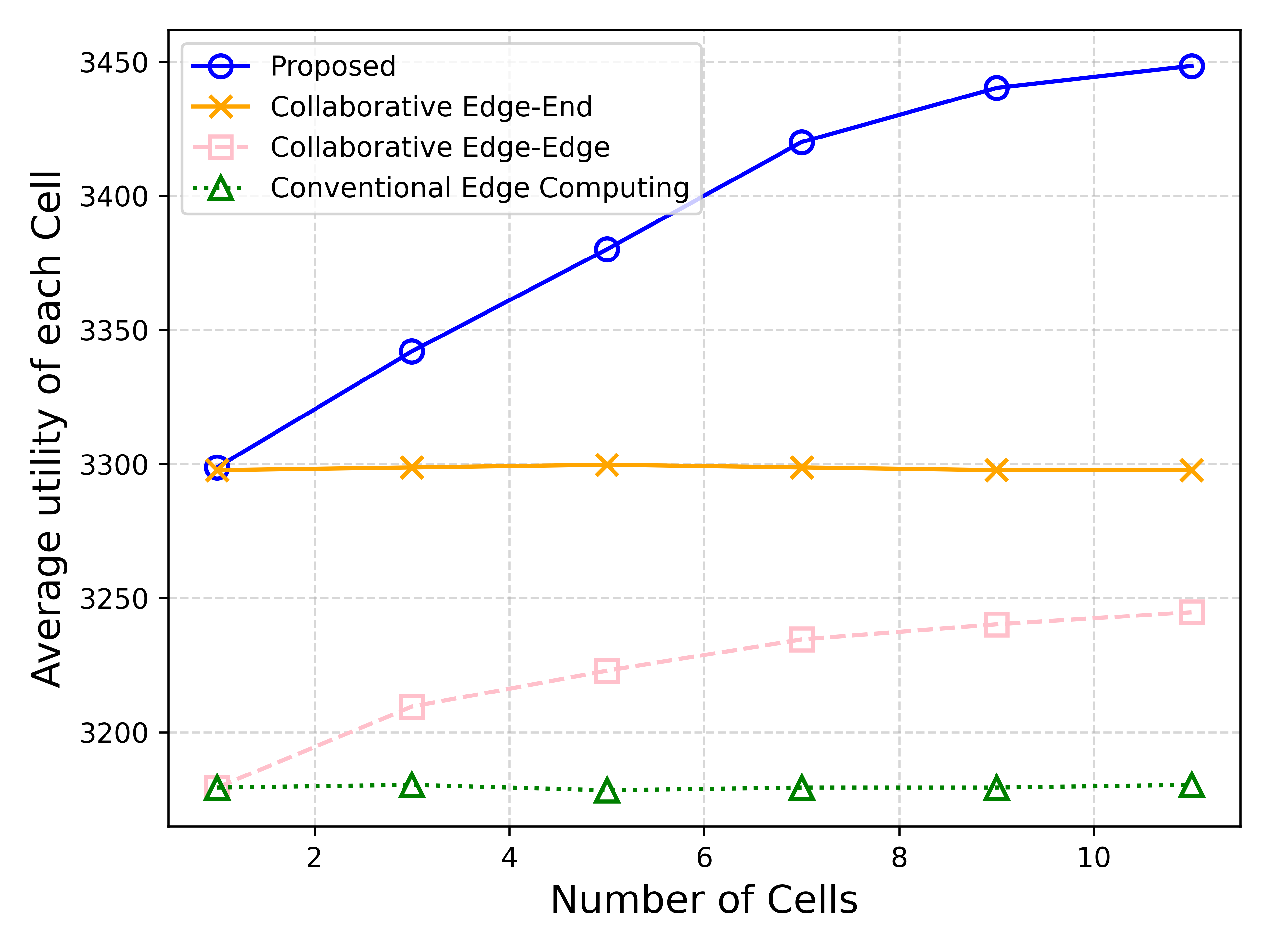}%
}
\subfloat[]{\includegraphics[width=0.33\textwidth]{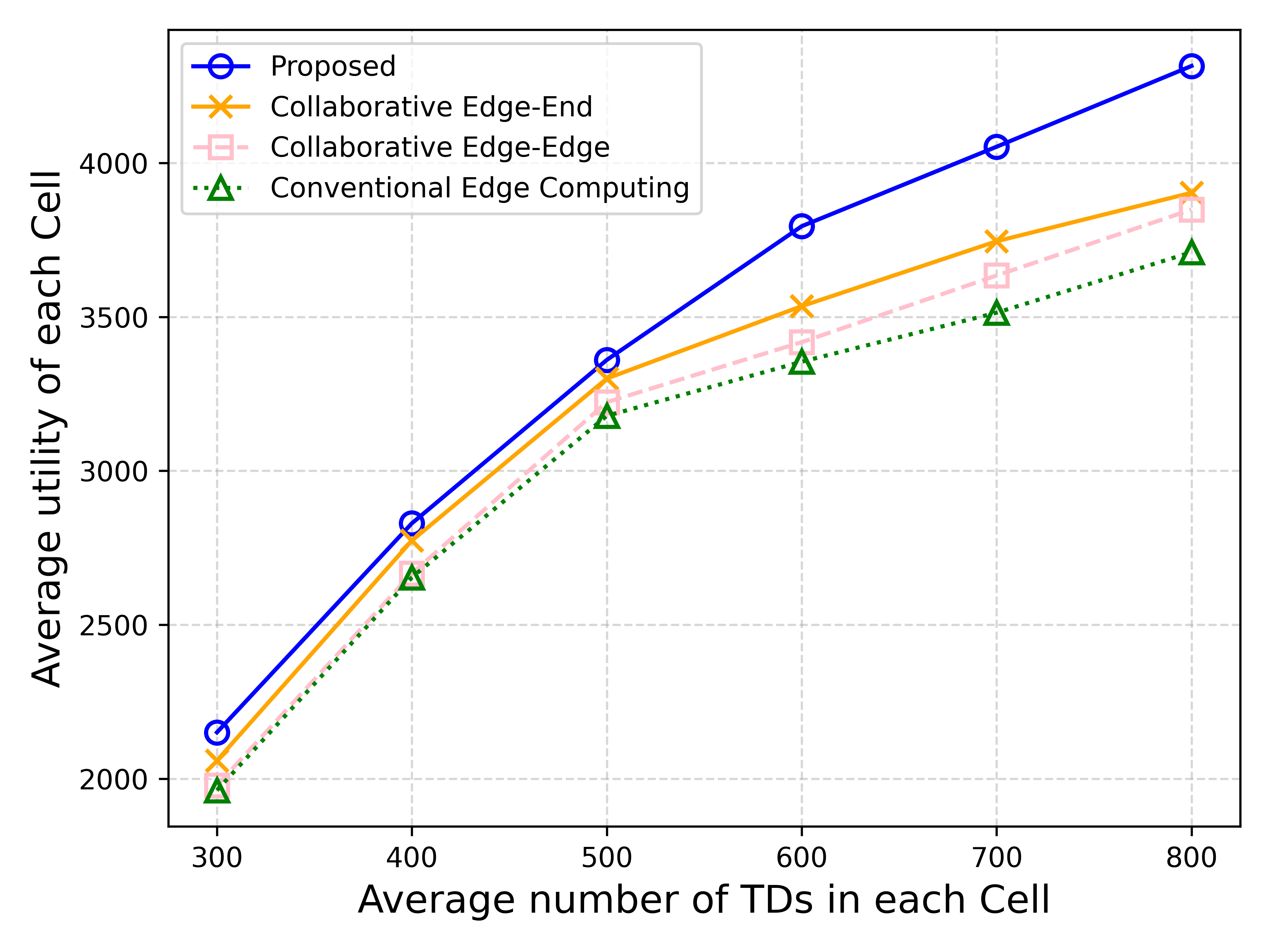}%
}
\subfloat[]{\includegraphics[width=0.33\textwidth]{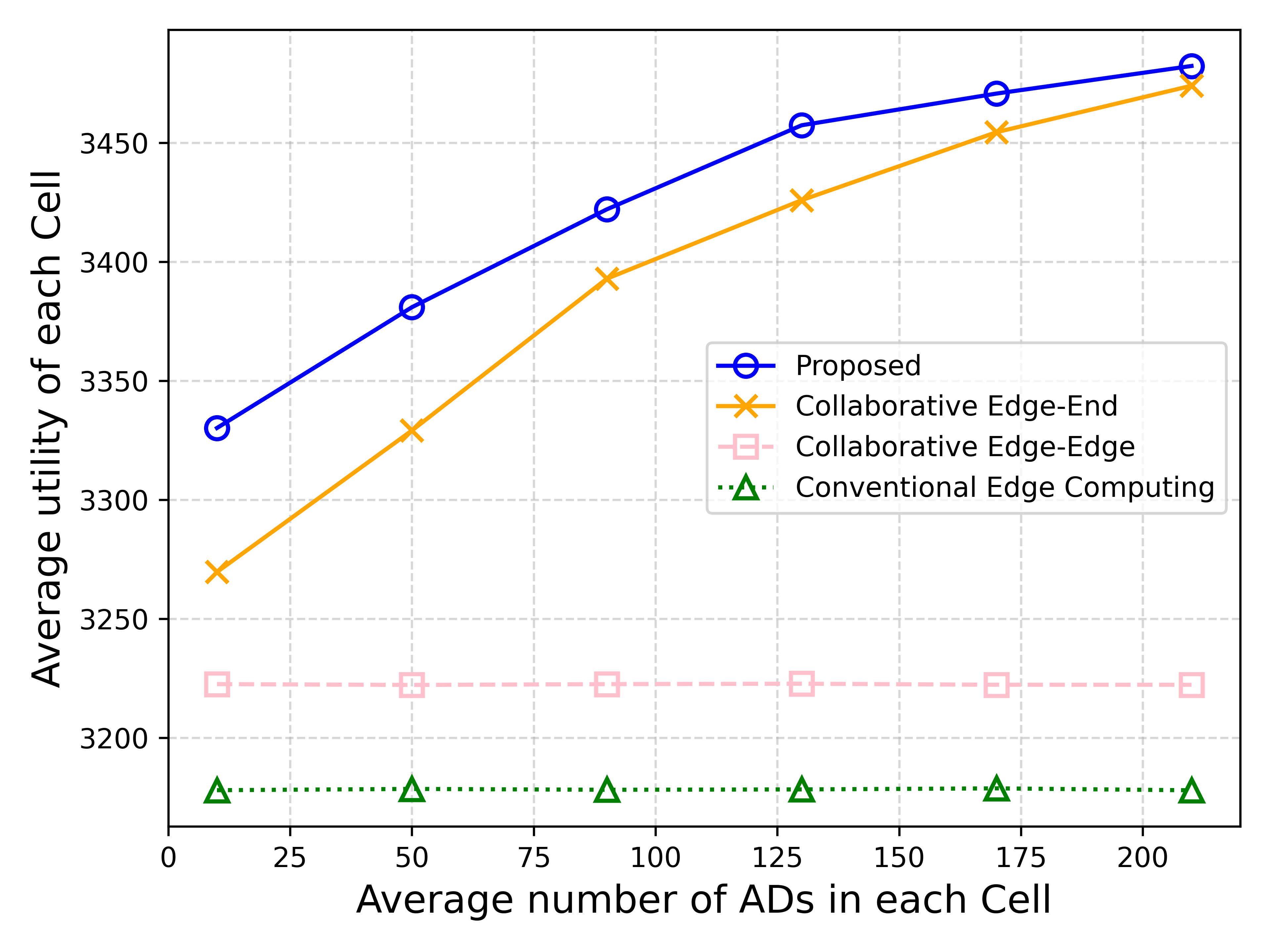}%
}
\caption{System utility of different frameworks. (a) Comparison of system utility with different numbers of cells. (b) Comparison of system utility with different numbers of TDs. (c) Comparison of system utility with different numbers of ADs.}
\label{System Utility}
\end{figure*}
\section{Performance Evaluation}

\subsection{Simulation Setting}\label{V-A}
In the simulation, we consider the scenario depicted in Fig. 1. Similar to \cite{34}, the transmit power of each IoT device is set to 2 W, the background noise power is -100 dBm, and the energy efficiency coefficient is $5 \times 10^{-27}$. According to \cite{31}, the unit energy cost is set to $1$ $\$$ and the task value ranges from $5$ $\$$ to $10$ $\$$. The main simulation parameters are summarized in Table \ref{tab2}. Unless otherwise specified, parameter values are set according to Table \ref{tab2}. All simulations were executed on a Python platform with an Intel Core i7-13650HX 4.9 GHz CPU and 16 GB of RAM.

We compare the performance of the proposed framework with the following benchmark schemes.

\begin{enumerate}
\item{\emph{Collaborative Edge-End Computing Strategy} \cite{40}: Auxiliary IoT devices serve as collaborators of edge servers for task processing but do not involve multiple ESs collaborating with each other.}
\item{\emph{Collaborative Edge Computing Strategy} \cite{41}: ESs collaborate to process tasks without assistance from ADs.}
\item{\emph{Double Auction-Based Strategy} \cite{38}: All ESs and ADs act as sellers, TDs act as buyers, and a trusted third-party platform acts as the auctioneer to organize the double auction.}
\item{\emph{Conventional Edge Computing Strategy}: ESs execute the arrived tasks directly without collaboration among ESs or assistance from ADs.}
\end{enumerate}

Notably, the double auction-based strategy involves new parameters (bids and offers of ESs and TDs) and is not comparable with the other schemes in terms of the objective function. Therefore, this scheme is only used to compare computational complexity, as shown in Fig. \ref{Running Time}.

In addition to comparing different frameworks, we also compared the algorithms used at each level of the proposed framework with the following baseline algorithms.
\begin{enumerate}
\item{\emph{Stackelberg Game}: A Stackelberg game is used in the first level of decision-making to determine the initial offloading decision and the payment fee for each TD.}
\item{\emph{Round-Robin Scheduling}: Round-robin scheduling is employed in the second level of decision-making to allocate tasks among ESs.}
\item{\emph{Vickrey Auction}: A Vickrey auction is employed in the third level of decision-making to incentivize ADs to assist ESs in processing tasks.}
\end{enumerate}

Notably, we modify only the algorithm of one level for comparison, while the algorithms in the other levels remain consistent with those in our proposed framework.

\subsection{Simulation Results}\label{V-B}
\emph{1) System Utility:} The system utility is defined as the average utility of all participants in each cell. Fig. \ref{System Utility}(a) shows the average system utility under various edge computing strategies with different numbers of cells. At $M=1$, the system utility is the same for both the proposed scheme and the collaborative edge-end computing strategy, as well as for the collaborative edge computing strategy and the conventional edge computing strategy, because there is only one cell. As the number of cells increases, the system utility of the collaborative edge-end computing strategy and the conventional edge computing strategy remain nearly constant, while that of the proposed scheme and the collaborative edge computing strategy gradually increases. This is because, as the number of cells increases, the utility gain from inter-cell collaboration becomes more prominent.

Fig. \ref{System Utility}(b) shows the variation in average system utility for different edge computing strategies with the mean number of TDs in each cell. In this experiment, $N_m, \forall m \in \mathcal{M}$ is uniformly distributed within a range with an interval length of 600. As the number of TDs increases, the system utility for all strategies increases because TDs can gain utility from processing their tasks locally. As the number of TDs increases, the number of tasks generated also increases, thus increasing the utility of processing these tasks. The proposed scheme achieves the highest system utility because it allows more TDs to offload their tasks. According to Section \ref{III-B}, TDs offload their tasks if and only if the utility gained from offloading is greater than from processing locally. Therefore, more TDs offloading their tasks results in higher system utility.

Fig. \ref{System Utility}(c) shows the variation in average system utility for different edge computing strategies with the mean number of ADs in each cell. Similarly, $K_m, \forall m \in \mathcal{M}$ is uniformly distributed within a range, with an interval length of 20. As the number of ADs increases, the system utility for the collaborative edge computing strategy and the conventional edge computing strategy remains nearly constant, while the system utility for the proposed scheme and the collaborative edge-end computing strategy gradually increases. This is because, as the number of ADs increases, the utility gain from ADs becomes more prominent, while the collaborative edge computing strategy and the conventional edge computing strategy do not utilize ADs.

\begin{figure*}[t]
\centering
\subfloat[]{\includegraphics[width=0.33\textwidth]{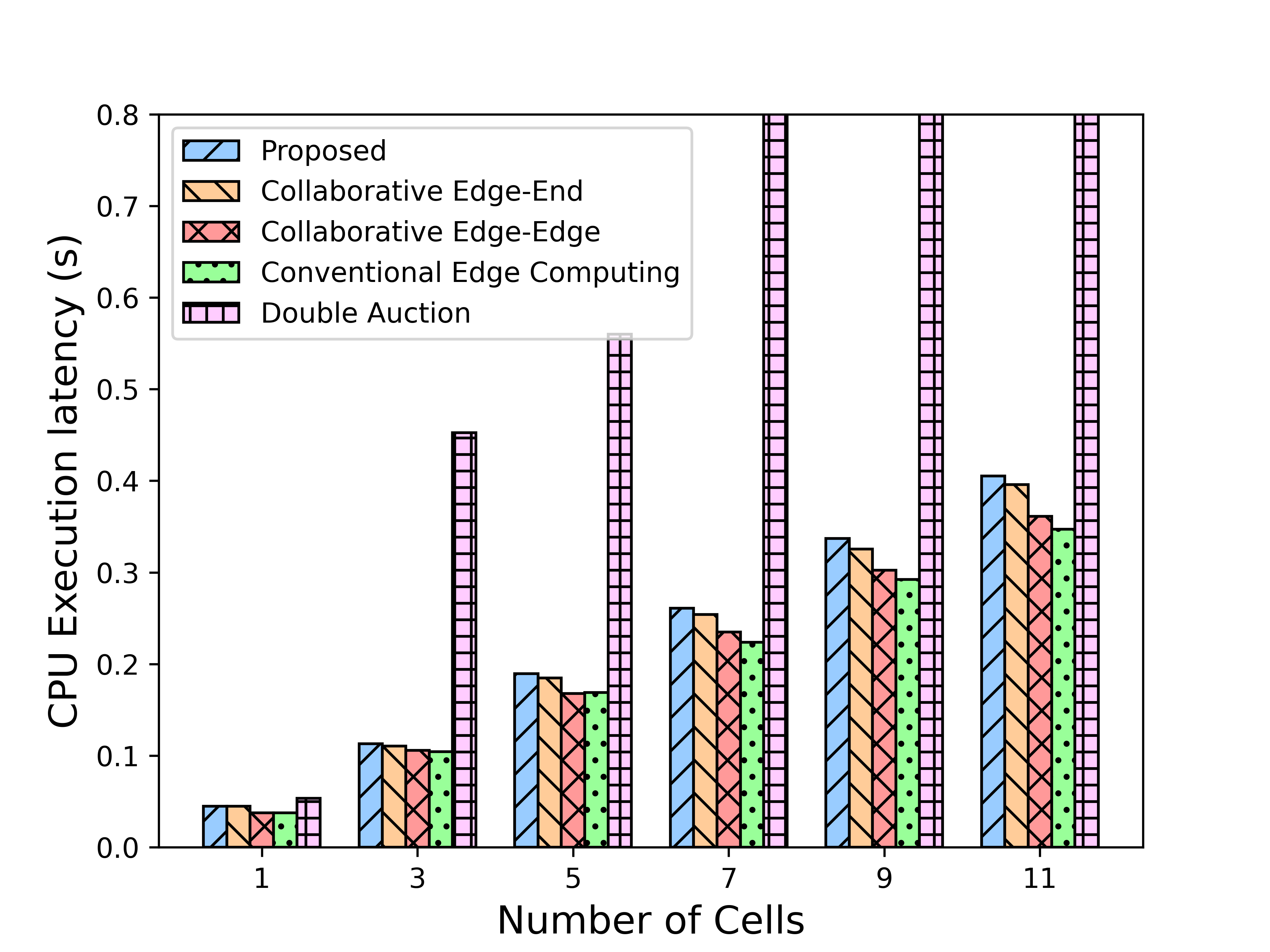}%
}
\subfloat[]{\includegraphics[width=0.33\textwidth]{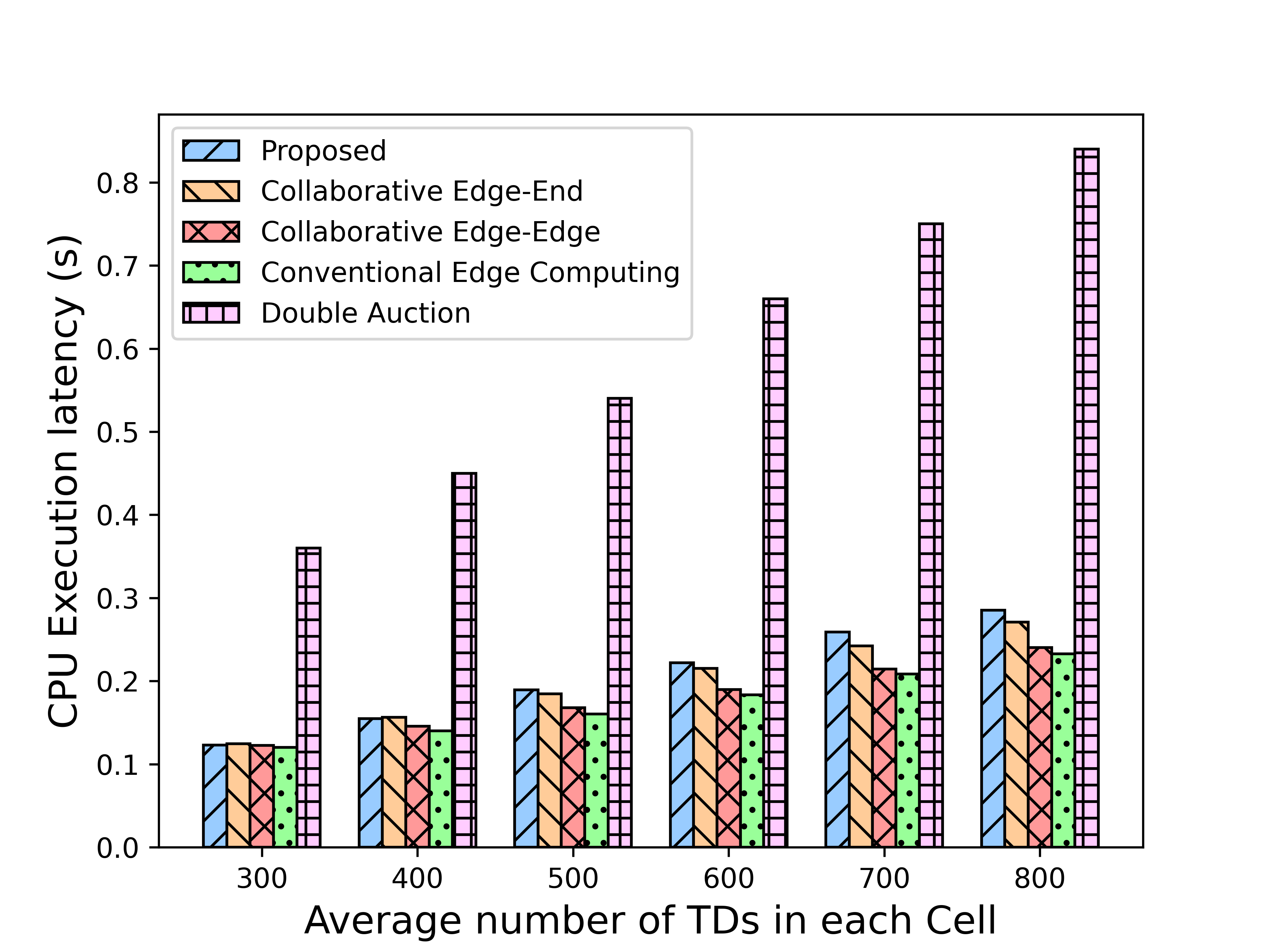}%
}
\subfloat[]{\includegraphics[width=0.33\textwidth]{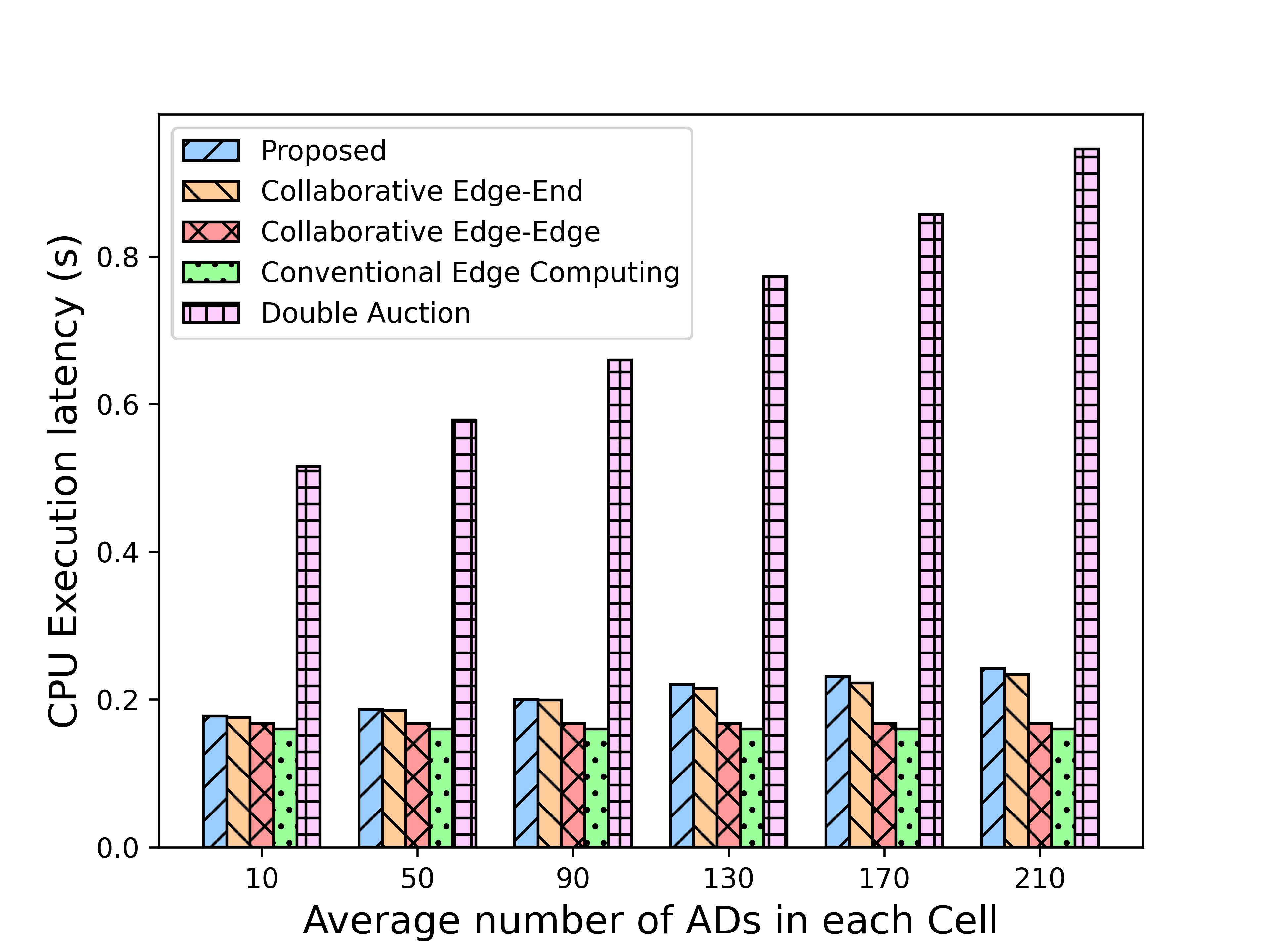}%
}

\caption{Average algorithm running time of different frameworks. (a) Comparison of algorithm running time with different numbers of cells. (b) Comparison of algorithm running time with different numbers of TDs. (c) Comparison of algorithm running time with different numbers of ADs.}
\label{Running Time}
\end{figure*}

\begin{figure*}[t]
\centering
\subfloat[]{\includegraphics[width=0.33\textwidth]{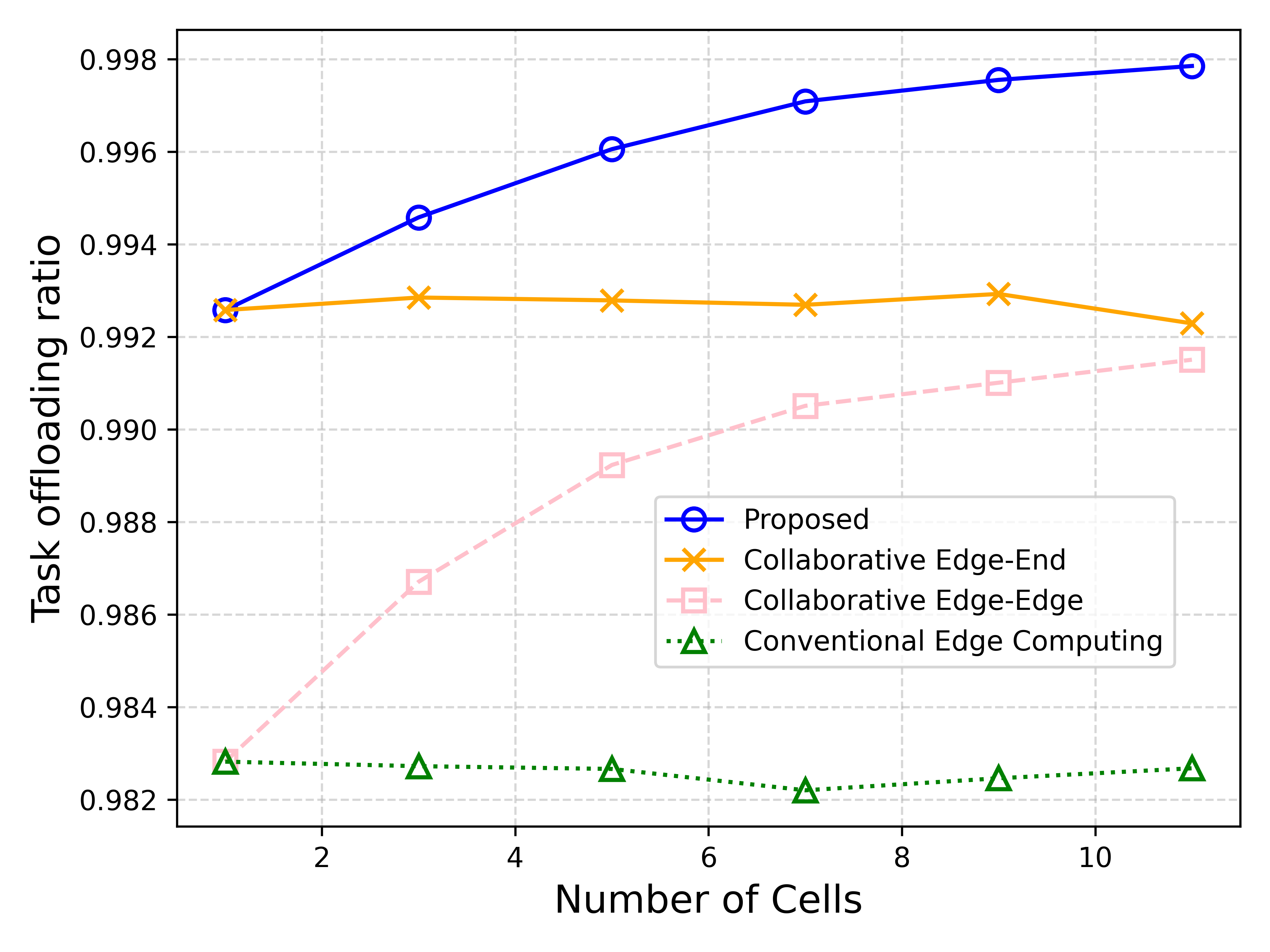}%
}
\subfloat[]{\includegraphics[width=0.33\textwidth]{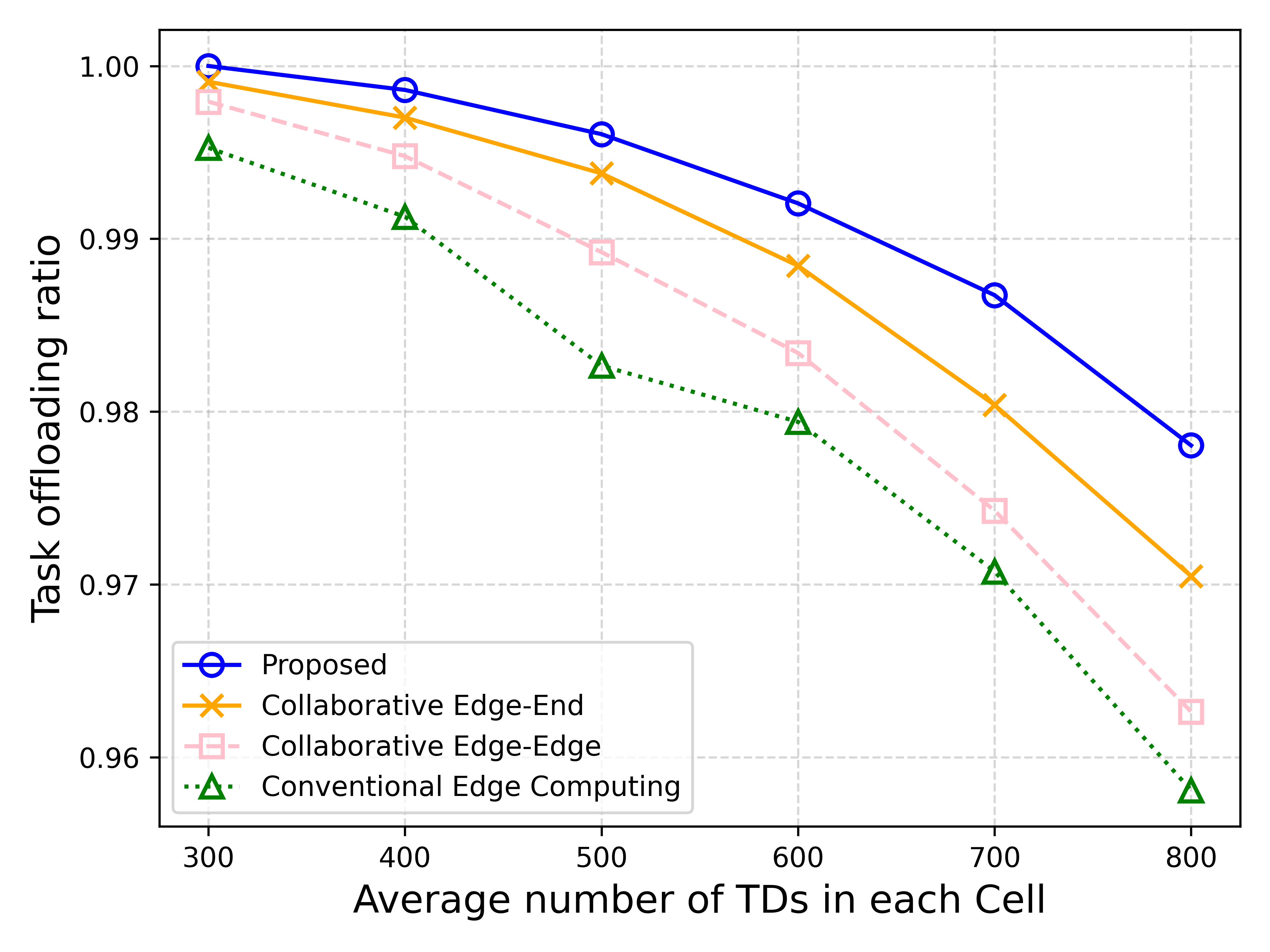}%
}
\subfloat[]{\includegraphics[width=0.33\textwidth]{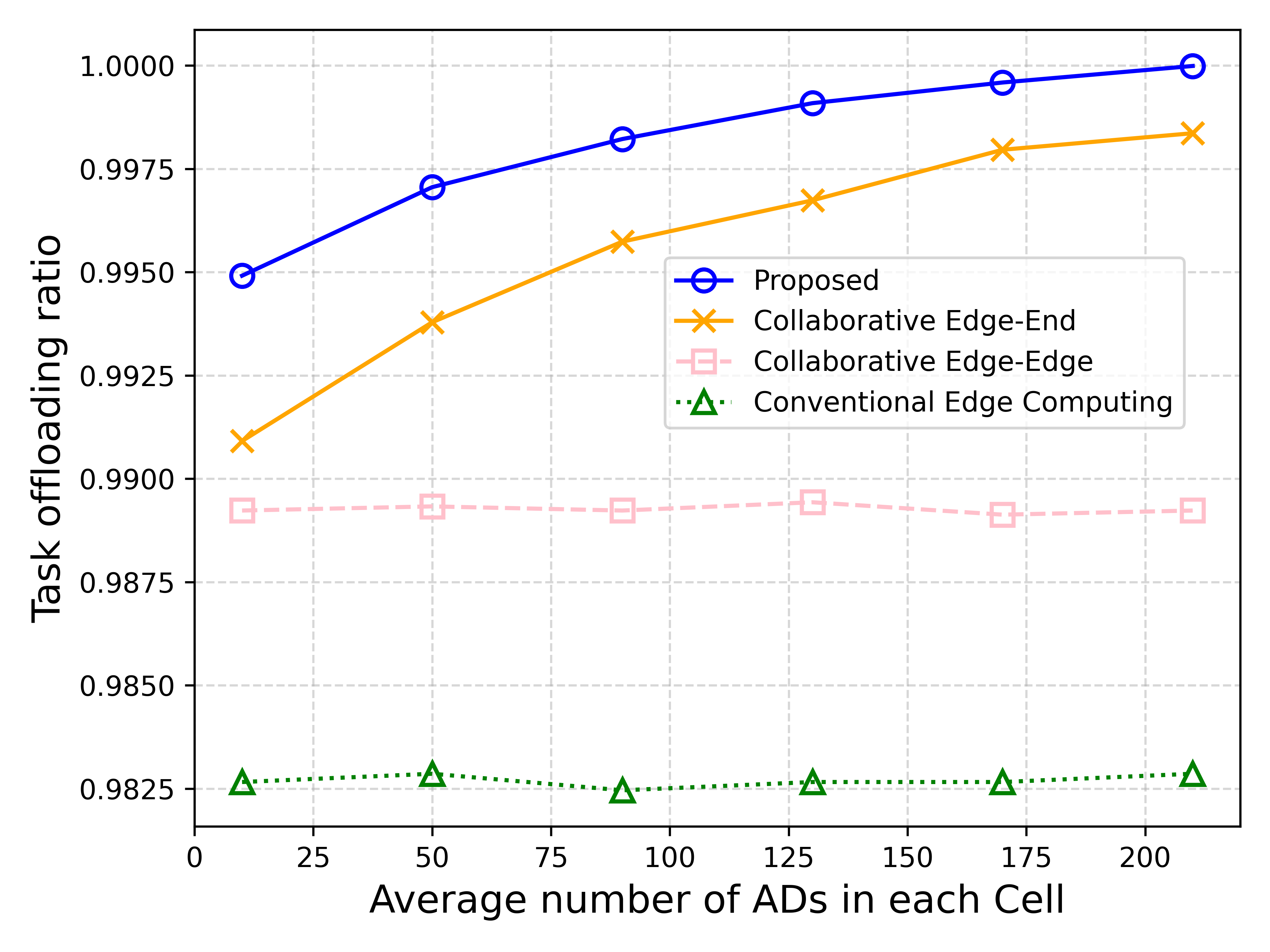}%
}

\caption{Average task offloading ratio of different frameworks. (a) Comparison of task offloading ratio with different numbers of cells. (b) Comparison of task offloading ratio with different numbers of TDs. (c) Comparison of task offloading ratio with different numbers of ADs.}
\label{Offloading Rate}
\end{figure*}

\emph{2) Algorithm Running Time:} Algorithm running time is defined as the average time taken for algorithm execution. Notably, in the simulation, we execute sub-algorithms serially that would be executed in parallel in a real scenario, so the algorithm's running time should be shorter in practice. Additionally, the parameter settings are consistent with the previous experiments. Fig. \ref{Running Time}(a) shows the average algorithm running times for different strategies versus the number of cells, indicating that the running times of all strategies increase with the number of cells. In real scenarios, due to parallel execution, the algorithm running time of the proposed strategy does not significantly increase with the number of cells. Fig. \ref{Running Time}(b) shows the variation in algorithm running times for different strategies with the mean number of TDs in each cell. As the mean number of TDs in each cell increases, the algorithm running times for all strategies grow. Fig. \ref{Running Time}(c) shows the variation in algorithm running times for different strategies with the mean number of ADs in each cell. As the number of ADs grows, the algorithm running times for the collaborative edge computing strategy and the conventional edge computing strategy remain nearly constant, while those of the proposed scheme and the collaborative edge-end computing strategy slowly increase.

From the three figures, it can be seen that the proposed strategy significantly reduces the algorithm execution latency compared to the double auction-based algorithm, mainly due to its hierarchical decision-making approach. Additionally, the algorithm running time of the proposed strategy is not sensitive to the number of TDs, ADs, or cells, making it suitable for online execution in large-scale IoT networks. Notably, the algorithm running time of the proposed scheme is longer than that of the collaborative edge-end strategy, collaborative edge computing strategy, and conventional edge computing strategy. This is because the latter three strategies only consider a subset of participants from the proposed scheme, effectively skipping one or two levels of decision-making, which leads to shorter running time. As illustrated in Fig. \ref{System Utility}, the proposed scheme achieves significant performance gains at the expense of only a slight increase in algorithm running time.

\emph{3) Average Task Offloading Ratio:} The task offloading ratio is defined as the ratio of the number of tasks eventually offloaded to the number of tasks decided to be offloaded in the first level of decision-making. Fig. \ref{Offloading Rate} shows the variation in task offloading ratio with different parameters for various strategies, with the experimental settings remaining the same. Fig. \ref{Offloading Rate}(a) shows the task offloading ratio versus the number of cells for different strategies. At $M=1$, the task offloading ratio is the same for both the proposed scheme and the collaborative edge-end computing strategy, as well as for the collaborative edge computing strategy and the conventional edge computing strategy. As the number of cells increases, the task offloading ratio for the conventional edge computing strategy and the collaborative edge-end computing strategy remains nearly constant, while that for the proposed scheme and the collaborative edge computing strategy gradually increases. This is because inter-cell collaboration allows the edge computing system to accommodate more offloaded tasks as the number of cells increases.

Fig. \ref{Offloading Rate}(b) shows the task offloading ratio versus the mean number of TDs in each cell for different strategies. It can be seen that the task offloading ratio decreases with the increasing number of TDs for all strategies, with the proposed strategy maintaining the highest ratio. Fig. \ref{Offloading Rate}(c) shows the task offloading ratio versus the mean number of ADs in each cell for different strategies. As the number of ADs increases, the task offloading ratio for the collaborative edge computing strategy and the conventional edge computing strategy remains nearly constant, while that for the proposed scheme and the collaborative edge-end computing strategy gradually increases. This is because, as the number of ADs increases, ADs can accommodate more offloaded tasks.

\begin{figure}[t]
\centering
\subfloat[]{\includegraphics[width=0.25\textwidth]{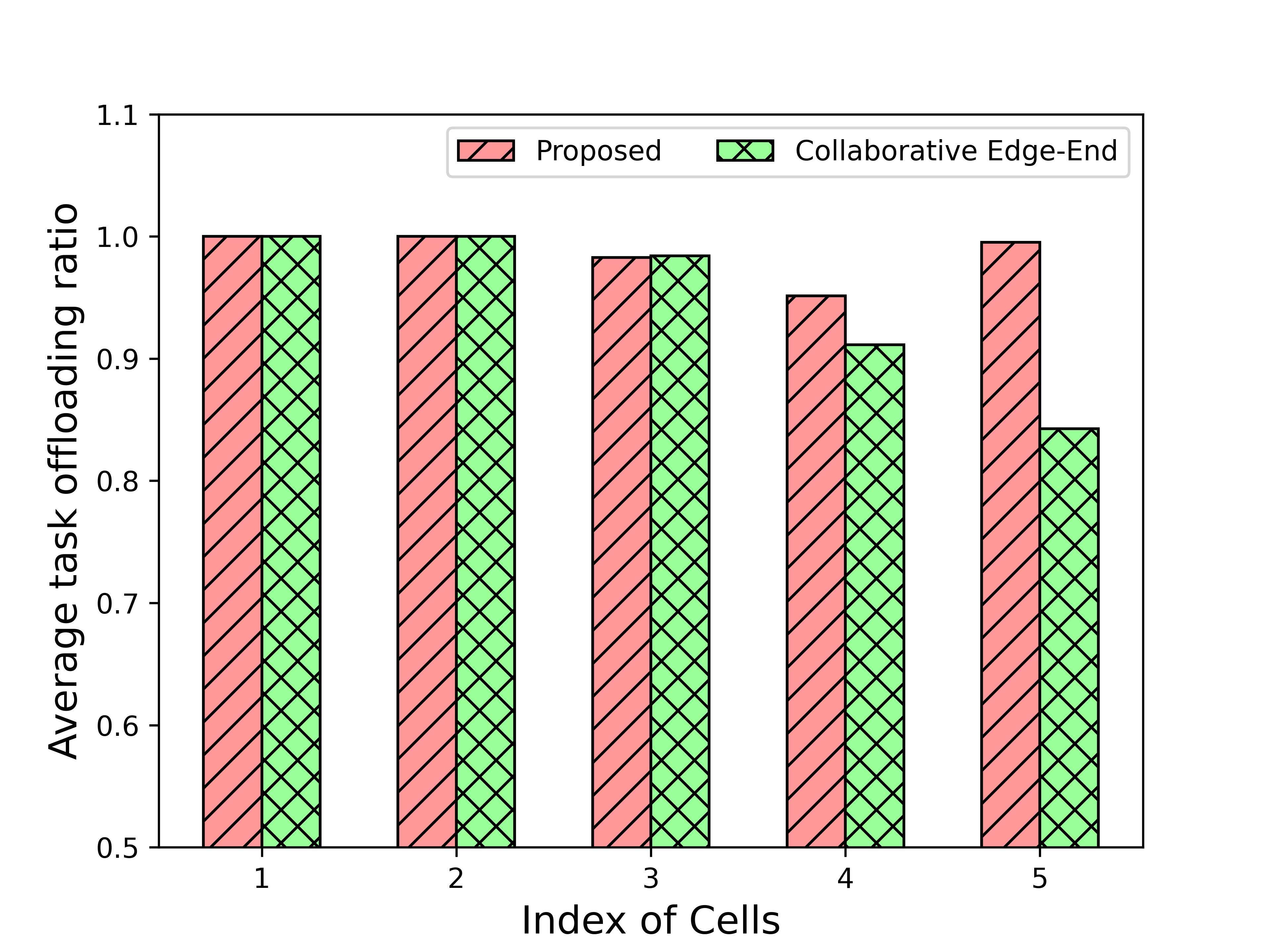}%
}
\subfloat[]{\includegraphics[width=0.25\textwidth]{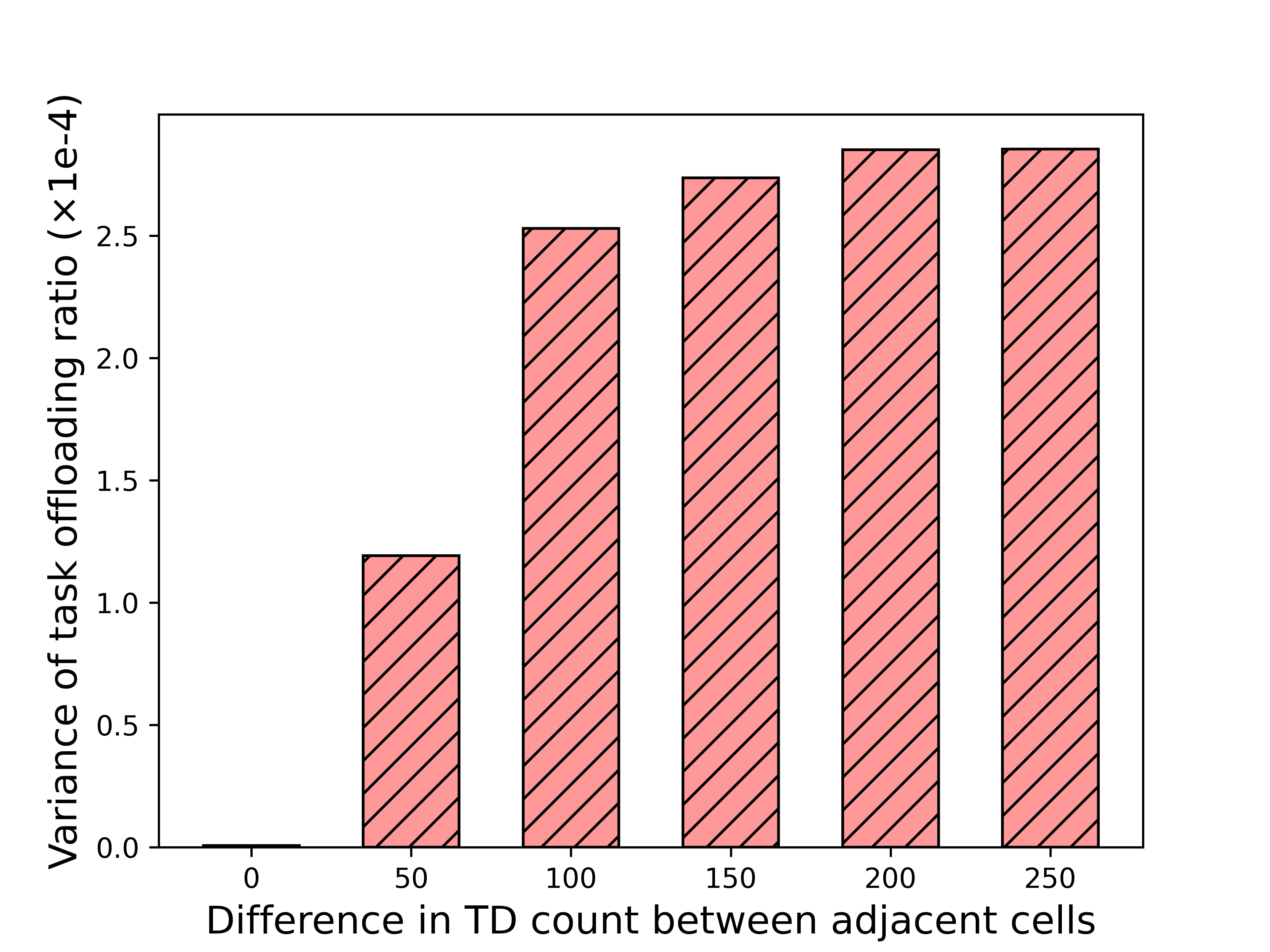}%
}
\caption{(a) The task offloading ratio in each cell with different frameworks. (b) The variance of task offloading ratio with different levels of task distribution non-uniformity for the proposed framework.}
\label{last}
\end{figure}

\begin{figure*}[t]
\centering
\subfloat[]{\includegraphics[width=0.33\textwidth]{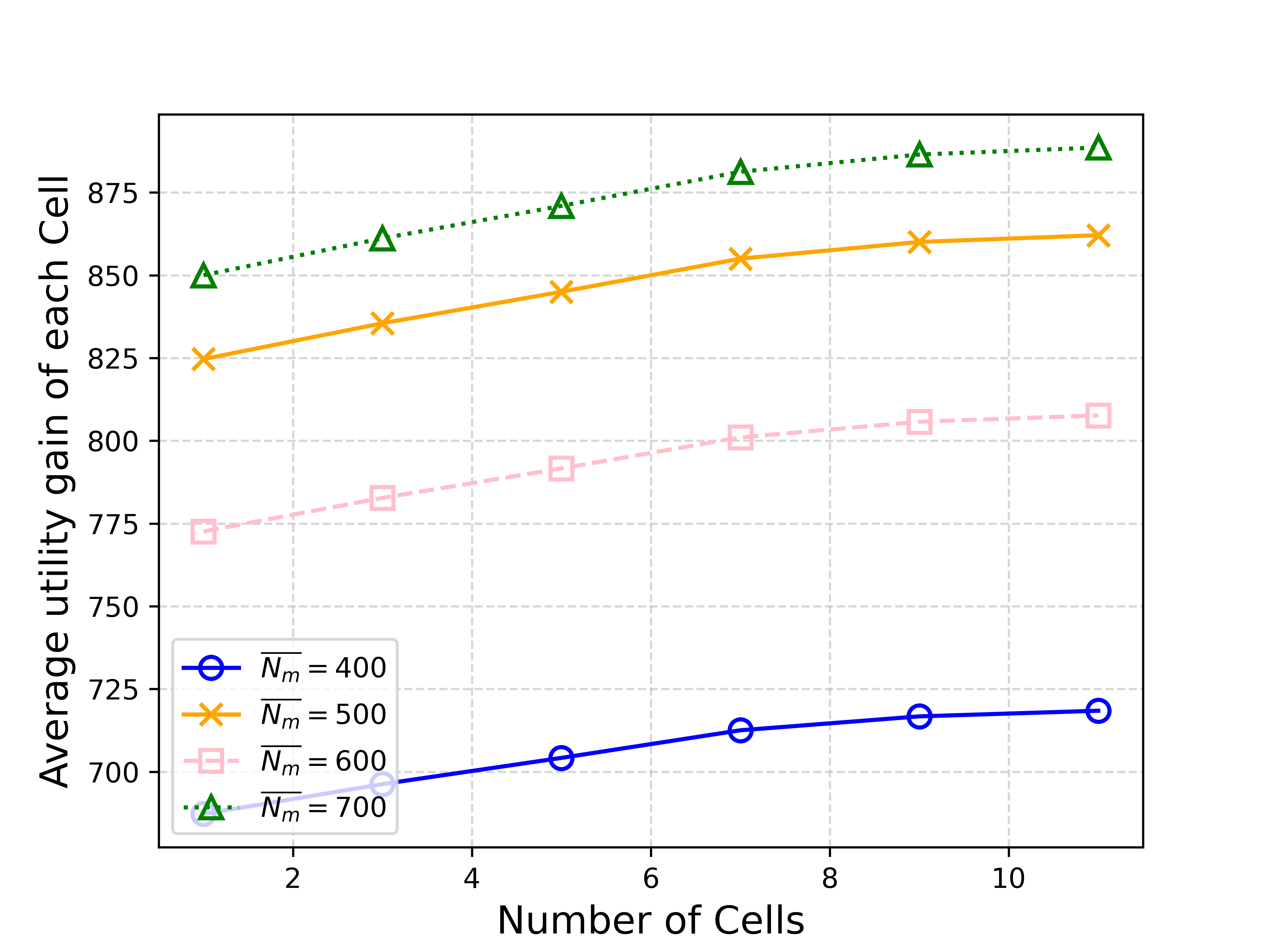}%
}
\subfloat[]{\includegraphics[width=0.33\textwidth]{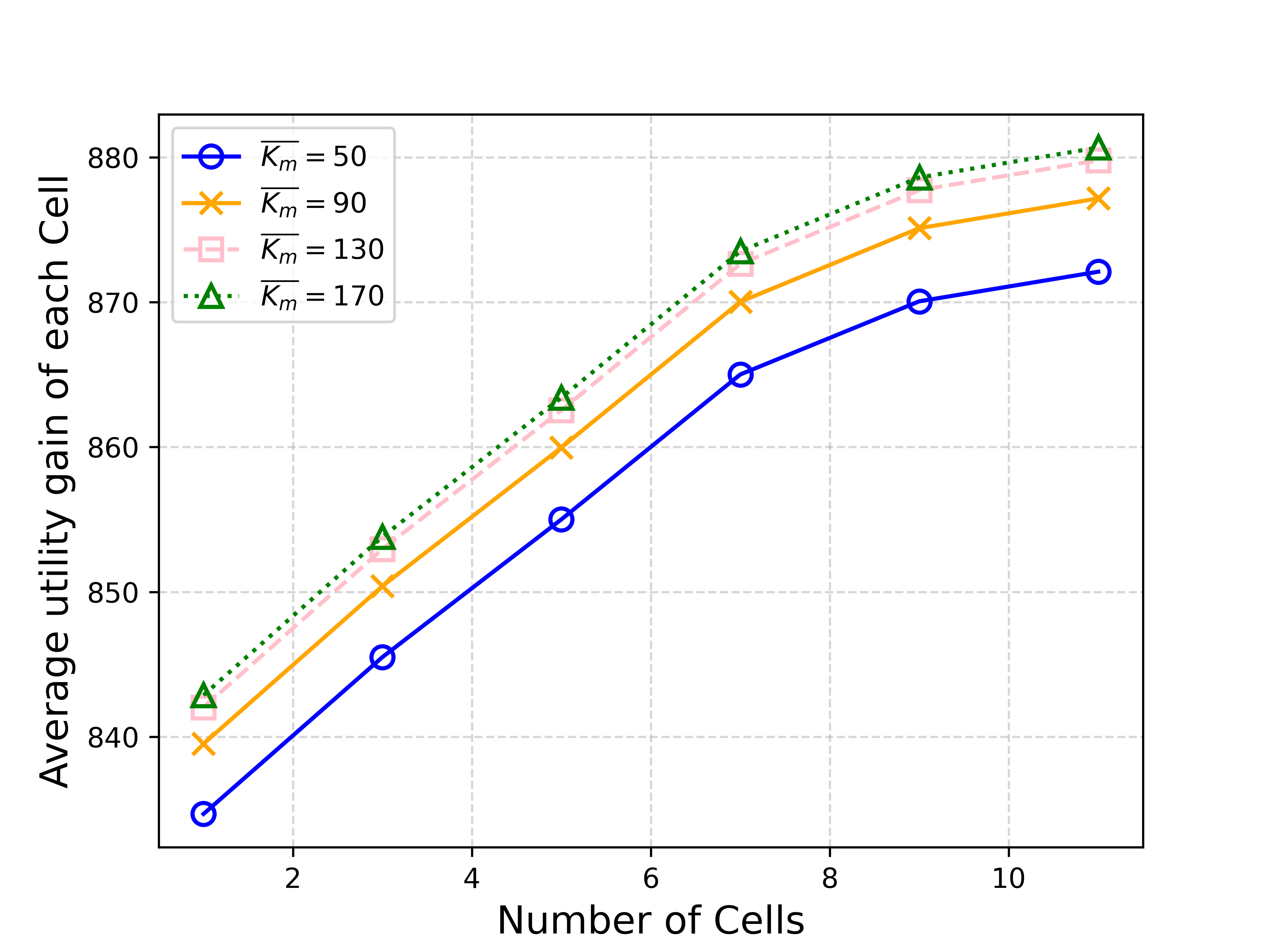}%
}
\subfloat[]{\includegraphics[width=0.33\textwidth]{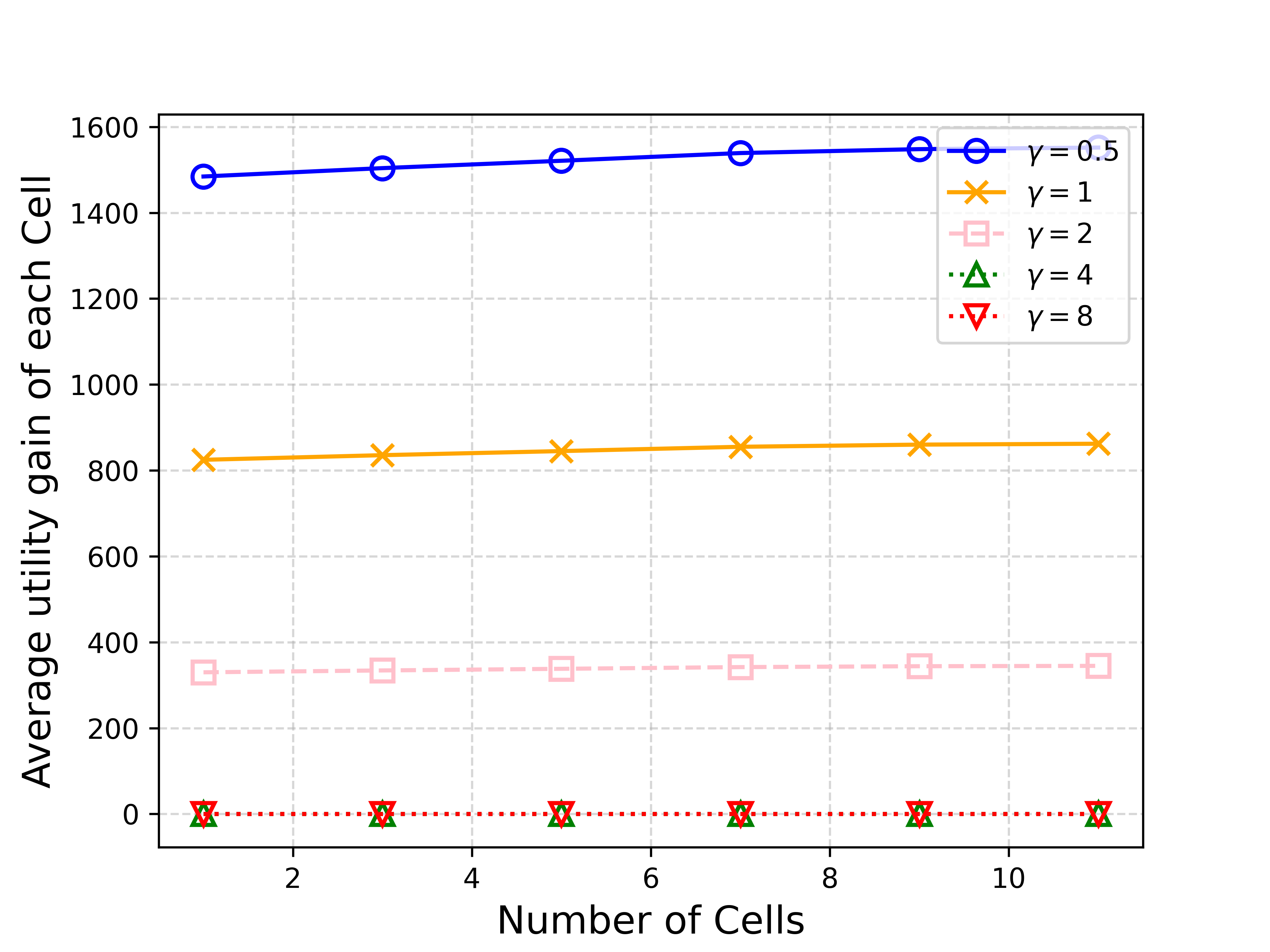}%
}

\caption{Impact of various system parameters on the utility gain achieved by the proposed scheme. (a) Comparison of utility gain with different average numbers of TDs per cell. (b) Comparison of utility gain with different average numbers of ADs per cell. (c) Comparison of utility gain with different unit energy costs.}
\label{utility gain}
\end{figure*}

\emph{4) Handle Non-Uniformly Distributed Tasks:} To evaluate the performance of the proposed framework with non-uniformly distributed tasks, we first compare it with the collaborative edge-end computing strategy, using the default values for the number of cells and the distribution of ADs from Table \ref{tab2}. Fig. \ref{last}(a) shows the task offloading ratio in each cell, with the number of TDs in the five cells set to 200, 350, 500, 650, and 800, respectively. It can be seen that the task offloading ratio does not differ much among cells under our proposed scheme. However, in the collaborative edge-end computing strategy, the task offloading ratio significantly decreases in the cells with more TDs. Fig. \ref{last}(b) shows the performance of the proposed scheme under different levels of task distribution non-uniformity. The horizontal axis represents the difference in the number of TDs between adjacent cells, with an average of 500 TDs across the five cells. It can be observed that although the variance of the task offloading ratio gradually increases as task distribution becomes more uneven, the variance remains at a relatively low level overall. In summary, the proposed scheme demonstrates good load-balancing capability.

\emph{5) Effect of Different Parameters on the Framework's Performance:} Fig. \ref{utility gain} illustrates the effect of various system parameters on the utility gain achieved by the proposed scheme. Here, utility gain refers to the increase in system utility relative to the scenario where all TDs adopt local execution mode. Fig. \ref{utility gain}(a) and Fig. \ref{utility gain}(b) depict how the system utility gain changes with an increasing number of cells, considering different average numbers of TDs and ADs per cell. As the number of cells increases, the utility gain from inter-cell collaboration grows, although the rate of growth progressively slows. Additionally, as the average number of TDs ($\overline{N_m}$) and ADs ($\overline{K_m}$) per cell increases, the system utility also rises, but the growth rate diminishes for different reasons. With an increase in the number of TDs, the available computing resources in the device-assisted edge computing network become inadequate to accommodate all TDs. When the number of ADs increases, the limited task processing demands of TDs cause more ADs to remain idle. Fig. \ref{utility gain}(c) shows how the system utility gain changes with varying unit energy costs as the number of cells increases. As $\gamma$ increases, the system utility gain decreases gradually, reaching zero when $\gamma \geq 4$. This occurs because as $\gamma$ increases, based on the expressions for $\alpha_{n_i^m}^{min}$ and $\alpha_{n_i^m}^{max}$ in (\ref{eq10}), $\alpha_{n_i^m}^{min}$ rises while $\alpha_{n_i^m}^{max}$ declines. When $\alpha_{n_i^m}^{min} \ge \alpha_{n_i^m}^{max}$, TD $n_i^m$ will execute the task locally, resulting in no utility gain for that task.

\begin{figure*}[t]
\centering
\subfloat[]{\includegraphics[width=0.33\textwidth]{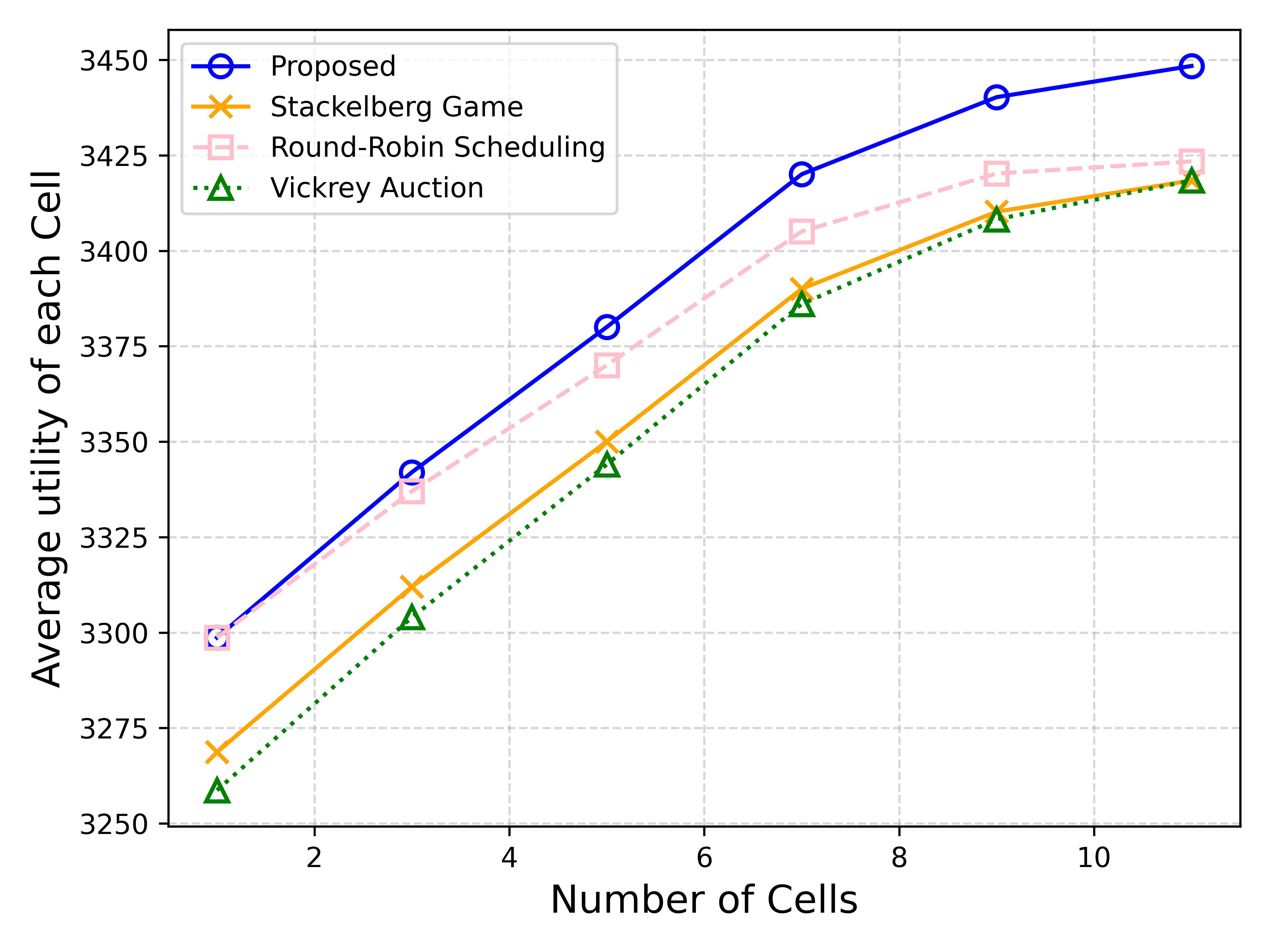}%
}
\subfloat[]{\includegraphics[width=0.33\textwidth]{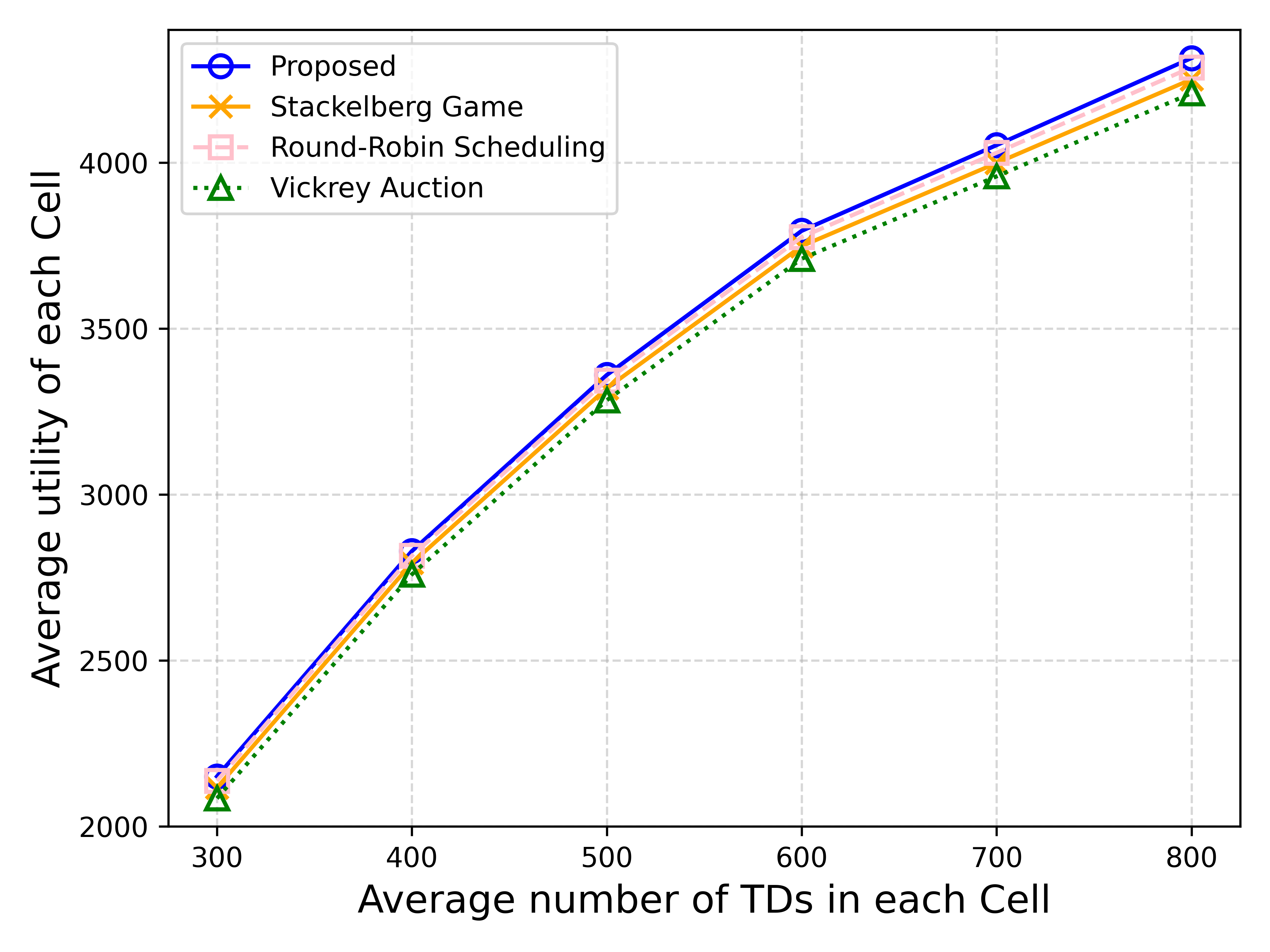}%
}
\subfloat[]{\includegraphics[width=0.33\textwidth]{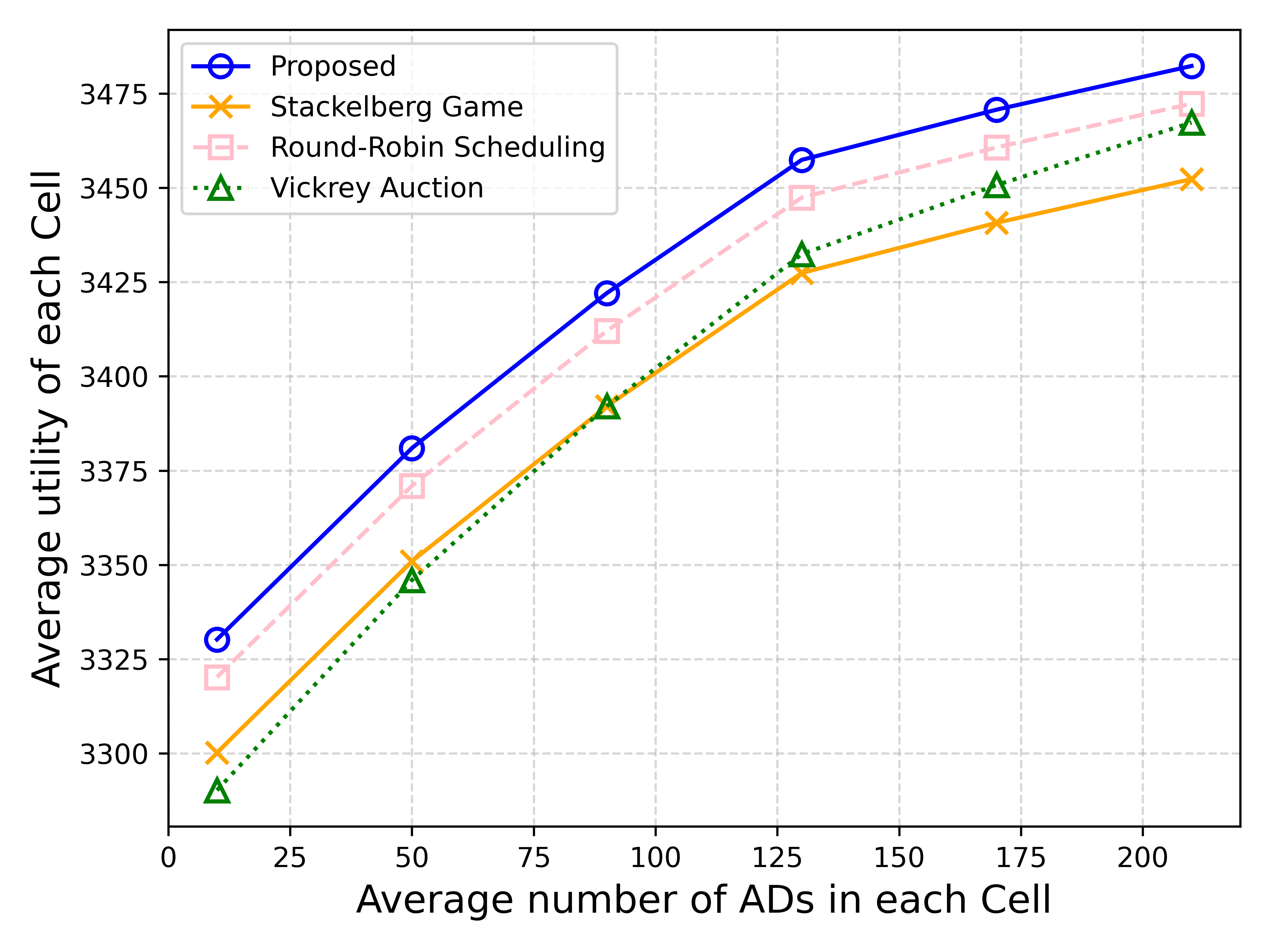}%
}
\caption{System utility of different algorithms. (a) Comparison of system utility with different numbers of cells. (b) Comparison of system utility with different numbers of TDs. (c) Comparison of system utility with different numbers of ADs.}
\label{different algorithms}
\end{figure*}

\emph{6) Algorithm Comparison:} Fig. \ref{different algorithms} compares the algorithms used in the proposed framework with the baseline algorithms, using an experimental setup consistent with Fig. \ref{System Utility}. The three sub-figures demonstrate that the algorithms used in the proposed framework are more aligned with the defined objective function, achieving higher system utility. This occurs because the Stackelberg game maximizes the utility of the leader (ESP) while neglecting the utility of the followers (TDs). Round-robin scheduling ensures fair resource allocation but does not aim to maximize utility. The Vickrey auction encourages buyers to bid honestly but ignores the lowest acceptable price for ESs' auction tasks, potentially reducing the utility of the ESP. Based on the experimental results and analyses, round-robin scheduling has the least effect on system utility. As shown in Fig. \ref{different algorithms}(a), the performance gap between round-robin scheduling and the proposed framework widens as the number of cells increases, while the gap between Vickrey auction and the proposed framework narrows. This trend is likely due to the increasing influence of the second decision-making level as the number of cells grows, which reduces the role of the third level in task-related decisions. Fig. \ref{different algorithms}(c) shows that as the average number of ADs per cell increases, the Vickrey auction curve surpasses the Stackelberg game, likely due to higher transaction prices for tasks in the third-level of decision-making as the average number of ADs per cell rises.

\section{Conclusion}
In this paper, we consider a multi-cell device-assisted edge computing system and propose an incentive-driven multi-level task scheduling framework that enables horizontal collaboration among edge servers and vertical collaboration between edge servers and auxiliary IoT devices. At the first level of decision-making, a bargaining game is used to model the interaction between TDs and ESs, ensuring fair and Pareto-optimal outcomes. At the second level of decision-making, we design a priority-based task scheduling algorithm to allocate tasks among ESs. At the third level of decision-making, the double auction mechanism is employed to incentivize ADs to assist ESs in processing tasks. Simulation results show that our proposed algorithm outperforms the benchmark strategies in terms of system utility, task offloading ratio, load balancing capability, and algorithm execution time. Future work involves edge servers from different edge service providers. This includes collaboration among edge servers controlled by the same provider and developing a reasonable profit-sharing mechanism to incentivize collaboration among different providers.

\bibliographystyle{IEEEtran}
\bibliography{IEEEabrv, references}

\begin{IEEEbiography}[{\includegraphics [width=1in,height=1.25in] {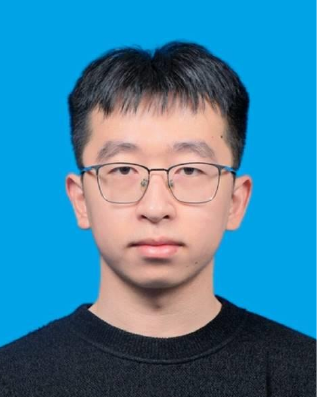}}] {Yang Li}
  received the B.S. degree in communication engineering from Beijing University of Posts and Telecommunications (BUPT), Beijing, China, in 2022. He is currently pursuing the Ph.D. degree with the Key Laboratory of Universal Wireless Communication, School of Information and Communication Engineering, BUPT. His research interests include device-assisted mobile edge networks, computing offloading and resource allocation.
\end{IEEEbiography}

\begin{IEEEbiography}[{\includegraphics [width=1in,height=1.25in] {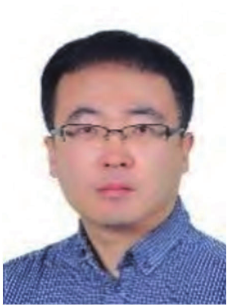}}] {Xing Zhang} 
  (M’10-SM’14) is currently a full professor with the School of Information and Communications Engineering, Beijing University of Posts and Telecommunications, China. His research interests are mainly in 5G/6G networks, satellite communications, edge intelligence, and Internet of Things. He has authored or coauthored five technical books and over 300 papers in top journals and international conferences and holds over 80 patents. He has received six Best Paper Awards in international conferences. He is a Senior Member of the IEEE and a member of CCF. He has served as a General Co-Chair of the third IEEE International Conference on Smart Data (SmartData-2017), as a TPC Co-Chair/TPC Member for a number of major international conferences.
\end{IEEEbiography}

\begin{IEEEbiography}[{\includegraphics [width=1in,height=1.35in] {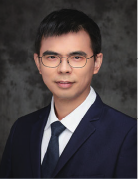}}] {Bo Lei}
  is now the deputy director of the Network Technology Research Institute of a director of future network research center of China telecom research institute. Bo Lei received his Master’s degree in Telecommunication Engineering from Beijing University of Posts and Telecommunications, Beijing, P. R. China, in 2006. His currently research interests include future network architecture, new network technology, computing power network and 5G application verification. Bo Lei now leads the future network research center focusing on future network. He is the first author of two technical books and has published more than 30 papers in top journals and international conferences, and filed more than 30 patents.
\end{IEEEbiography}

\begin{IEEEbiography}[{\includegraphics [width=1in,height=1.35in] {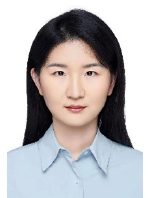}}] {Qianying Zhao}
  is now an engineer of future network research center of China telecom research institute. Qianying Zhao received her Master’s degree in Sant'Anna School of Advanced Studies, Pisa, Italy, in 2018. Her currently research interests include computing power network, edge computing. She is the author of two technical books and has published 7 papers in journals.
\end{IEEEbiography}

\begin{IEEEbiography}[{\includegraphics [width=1in,height=1.35in] {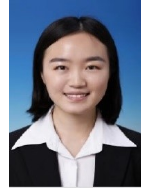}}] {Min Wei}
  is now an engineer of future network research center of China telecom research institute.  She received her Master’s degree in Beijing University of Posts and Telecommunications, Beijing, P. R. China, in 2015. Her currently research interest is computing power network.
\end{IEEEbiography}

\begin{IEEEbiography}[{\includegraphics [width=1in,height=1.25in] {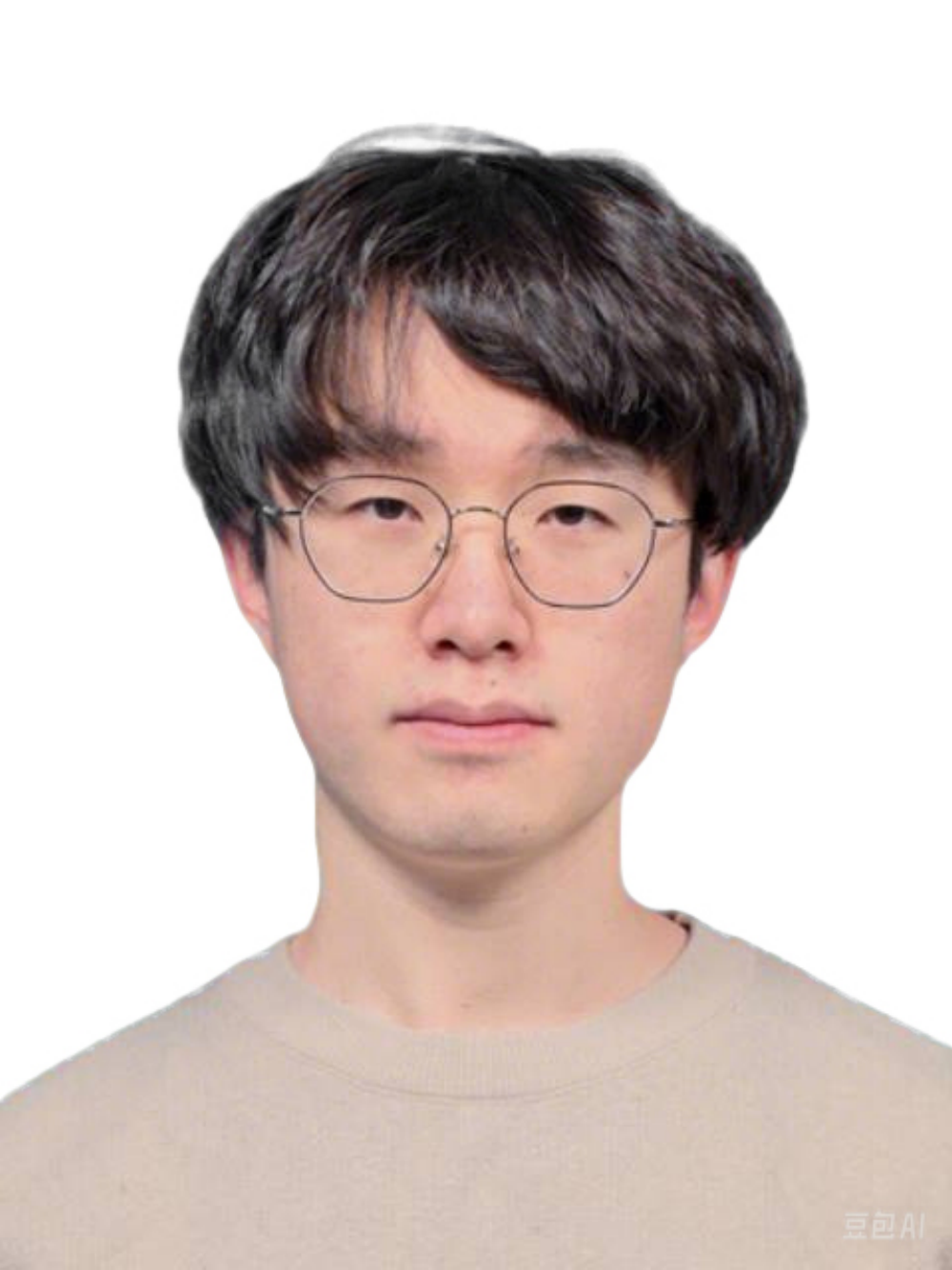}}] {Zheyan Qu}   
  received his B.S. degree in information engineering at the Beijing University of Posts and Telecommunications (BUPT), Beijing, China, in 2022. He is currently pursuing the M.S. degree with the Key Laboratory of Universal Wireless Communication, School of Information and Communication Engineering, BUPT. His research interests include edge computing, natural language processing, and large language models.
\end{IEEEbiography}

\begin{IEEEbiography}[{\includegraphics [width=1in,height=1.35in] {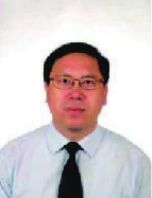}}] {Wenbo Wang}
  received the B.S., M.S., and Ph.D. degrees from BUPT in 1986, 1989, and 1992, respectively. He is currently a Professor with the School of Information and Communications Engineering, and the Executive Vice Dean of the Graduate School, Beijing University of Posts and Telecommunications. He is currently the Assistant Director with the Key Laboratory of Universal Wireless Communication, Ministry of Education. He has authored over 200 journal and international conference papers, and six books. His current research interests include radio transmission technology, wireless network theory, and software radio technology.
\end{IEEEbiography}

\end{document}